# Earthquake precursors in the light of peroxy defects theory:

# critical review of systematic observations

**Friedemann Freund**
GeoCosmo Science Center, NASA Ames Research Park, Moffett Field, CA 94035, USA
Friedemann.T.Freund@nasa.gov

**Guy Ouillon**
Lithophyse, Nice, France and ETH Zurich, Switzerland
Ouillon@aol.com

**John Scoville**
GeoCosmo Science Center, NASA Ames Research Park, Moffett Field, CA 94035, USA
john.c.scoville@nasa.gov

**Didier Sornette**
ETH Zurich (Swiss Federal Institute of Technology in Zurich)
Department of Management, Technology and Economics (D-MTEC) and
Department of Earth Sciences and Department of Physics
Switzerland
dsornette@ethz.ch
www.er.ethz.ch

**Abstract**: Forecasting earthquakes implies that there are time-varying processes, which depend on the changing conditions deep in the Earth's crust prior to major seismic activity. These processes may be linearly or non-linearly correlated. In seismology, the research has traditionally been centered on mechanical variables, including precursory ground deformation (revealing the build-up of stress deep below) and on prior seismic events (past earthquakes may be related to or even trigger future earthquakes). Since the results have been less than convincing, there is a general consensus in the geoscience community that earthquake forecasting on time scales comparable to meteorological forecasts are still quite far in the future, if ever attainable.

The starting point of the present review is to acknowledge that there are innumerable reports of other types of precursory phenomena ranging from emission of electromagnetic waves from ultralow frequency (ULF) to visible (VIS) and near-infrared (NIR) light, electric field and magnetic field anomalies of various kinds (see below), all the way to unusual animal behavior, which has been reported again and again. These precursory signals are intermittent and seem not to occur systematically before every major earthquake and reports on pre-earthquake signals are not widely accepted by the geoscience community at large because no one could explain their origins. In addition, the diversity of the signals makes them look disparate and unrelated, hampering any progress.

We review a credible, unifying theory for a solid-state mechanism, which is based on decades of research bridging semi-conductor physics, chemistry and rock physics. This theory, which we refer to as the "peroxy defect theory", is capable of providing explanations for the multitude of reported pre-earthquake phenomena. A synthesis has emerged that all pre-earthquake phenomena could trace back to one fundamental physical process: the activation of electronic charges (electrons and positive holes) in rocks subjected to ever-increasing tectonic stresses prior to any major seismic activity, via the rupture of peroxy bonds. The holes are unusual inasmuch as they are able to flow out of the stressed rock volume, into and through surrounding unstressed or less stressed rock, forming electric currents, traveling fast and far, and causing a wide range of physical and chemical follow-on processes along the way, which can be measured, ranging from electrical ground potentials, stimulated infrared emission, massive air ionization, to increased levels of ozone and toxic levels of carbon monoxide (CO).

In the second part of the review, we critically examine satellite and ground station data, recorded before past large earthquakes, as they have been claimed to provide evidence that precursory signals tend to become measurable days, sometimes weeks before the disasters. We review some of the various phenomena that can be directly predicted by the peroxy defect theory , namely, radon gas emanations, corona discharges, thermal infrared emissions, air ionization, ion and electron content in the ionosphere, and electro-magnetic anomalies.

Our analysis demonstrates the need for further systematic investigations, in particular with strong continuous statistical testing of the relevance and confidence of the precursors. Only then, the scientific community will be able to assess and improve the performance of earthquake forecasts.

# Table of contents

# I-Introduction

## 1.1 The cost of earthquakes

Large earthquakes are, by far, the deadliest of all natural disasters, claiming an average of 60,000 lives a year, featuring gigantic fluctuations (e.g. 80,000 victims from 1994 to 2004, and 780,000 from 2001 to 2010; see e.g. Knopoff and Sornette, 1995, for the size distribution of death tolls related to seismic events), which partly mirrors the highly intermittent distribution of seismicity in space, time, and magnitude. On more economic grounds, such disasters are also causing colossal property and industrial damage, with that of the 1989 Loma Prieta earthquake in California alone estimated at $6 billions (over €4 billions), the 1995 Kobe event in Japan estimated at $200 billions (€150 billions), while the 2011 Tohoku earthquake followed by its great tsunami already stands with much higher losses, with costs continuing to rise with the on-going management of the Fukushima nuclear disaster, which is likely to last several decades. Taking into account lost productivity, lost income, lost tax revenue, as well as the cost of rebuilding all infrastructures, the economic impact of a magnitude 7 or larger earthquake is expected to exceed €100 billions. The situation can only become more acute with the on-going growth and concentration of human populations in urban centers often found in seismic regions.

In this backdrop, if it were possible to warn of an impending major earthquake days or even weeks in advance, damage to industry, transportation and the power grid could be significantly reduced by taking appropriate mitigation measures. Numerous lives could be saved. By allowing recovery to begin sooner, the post-disaster restoration phase could also be made more cost-effective and more efficient.

## 1.2 Seismology: a very short historical introduction

Instruments to detect earthquakes appeared quite early in History, such as the Chinese seismoscope invented by Zhang Heng in 132 CE. However, those prototypes were unable to record and store any signal, hence did not allow their contemporaries (nor us) to provide precise location, origin time or energy estimation for the observed events. For a long period of time (i.e. until the late XIX$^{th}$ century), human beings and edifices have played the role of rudimentary seismographs (and accelerometers): through the numerous reports that one may find in historical archives, historians and scientists are able to list a catalog of large past events in inhabited areas. The amount of observed damage and perceptions of the people allow one to draw earthquake intensity maps, from which one can deduce estimations of the location of the epicenter and, according to quite dispersed empirical laws, its magnitude and depth. Even if genuine seismological networks appeared only during the XX$^{th}$ century, allowing a quantitative description of seismic ruptures, some important aspects of modern seismology appear to be rooted in philosophical reflections (about what we now coin as *seismic risk*) during the French Renaissance, in the wake of the great Lisbon earthquake in 1755, when a famous controversy opposed French philosophers François-Marie Arouet (better known as Voltaire) and Jean-Jacques Rousseau. The former (Voltaire, 1756) possessed what we would now qualify as a more hazard-oriented state of mind (claiming that such disasters occur randomly), while the latter (Rousseau, 1756) certainly formulated the very first ideas about risk, i.e. that man could protect himself from such calamities by coming up with better-thought building and urbanization policies. This hazard and risk dichotomy and terminology still defines nowadays the two main classes of approaches to natural catastrophes in the academic and engineering worlds, respectively.

## 1.3 Seismology as a science

On the side of hard science, a first major advance has been the model published by Reid (1910) who, for the first time, bridged different physical phenomena known both in nature and in the laboratory: earthquakes, faulting, and friction. Reid's elastic rebound theory not only allowed to explain the phenomenology of earthquakes, it also provided huge hopes in the possibility of predicting the large, devastating events that were threatening California and many other places worldwide. Reid viewed a seismic prone region as containing an isolated fault considered as a perfectly periodic relaxation oscillator with threshold dynamics, with each relaxation event constituting a large characteristic earthquake. His simple model has indeed more or less survived until now, with many refinements, for instance through the stress transfer approach to earthquake forecasting based on a deterministic analysis of fault-fault interactions (King et al., 1994; Nandan et al., 2016). While this approach can explain *a posteriori* some space-time features of some observed aftershock sequences following some large events (see for instance Bhloscaidh and McCloskey, 2014), it never provided any reliable prediction of similar earthquake sequences.

On the other hand, earthquake data began to accumulate at the turn of $XIX^{th}$ to $XX^{th}$ century, allowing the use of purely statistical descriptions (with many of them quite abusively and unfortunately ending up as so-called *laws* in the earthquake science terminology). The first of these laws quantifies the temporal rate of earthquakes following a large main event (and known as the Omori law; see Omori, 1894), and has been since similarly extended to space. The second one is the famous Gutenberg-Richter law, which is nothing but the power-law shaped energy distribution of observed events (Gutenberg and Richter, 1956). Combining these two robust statistical features (complemented by few other ones, like the productivity law (Helmstetter, 2003) quantifying the number of aftershocks directly triggered by a given event as a function of its magnitude) allowed the new branch of statistical seismology (and forecasting) to emerge.

Forecasting in seismology can be subdivided into two branches: (i) time-independent forecasting (which simply approximates seismicity as a constant-rate Poisson process in a given region, an approach initially popularized by Cornell, 1968), and (ii) time-dependent forecasting, which models seismicity as a linear superposition of generations of events triggered by all past events, allowing to forecast the seismicity rate at future times (see for instance Ogata, 2011). Time-independent and time-dependent approaches are now evaluated and compared within a well-defined framework which allows one to test their mutual performances (see the Collaboratory for the Study of Earthquake Predictability, http://www.cseptesting.org). All these techniques continue to be improved and now feature impressive degrees of sophistication. However, they only allow to define probabilities of occurrence within a time scale of a few years at least. This is clearly insufficient as one goal of earthquake prediction is to define efficient short-term mitigation strategies such as temporary population evacuation or critical infrastructures shutdown. This severe drawback is mainly due to the complete lack of constraint on the magnitudes of upcoming events: the time-varying poissonian rate of occurrence of the next event can be modelled with some reasonable accurracy, but its magnitude is simply modelled by a random sampling of the time-invariant Gutenberg-Richter law (Helmstetter and Sornette, 2003). This is why such approaches do not define any prediction methodology *per se*.

Another branch independently emerged in the 1990s, pushing forward analogies between the many power-laws observed in the phenomenology of earthquakes and the dynamics of critical phase transitions (see e.g. Bowman et al., 1998). This approach led to consider the time of large events as critical points (or at least finite-time singularities), thus suggesting some predictability especially for large earthquakes, implying that the larger the event, the better their predictability (Huang et al., 1998). While this concept seemed to apply well in the laboratory or small-scale systems such as mines (Ouillon and Sornette, 2000) and under stress-controlled conditions, its applicability to large scale, strain-controlled systems such as tectonic plates is still debated (Mignan, 2011). Other approaches based on pattern recognition are also in use but do not provide really convincing results yet as they have not been thoroughly tested in real time. The Russian team around Keilis-Borok and Kossobokov has developed a rigorous testing framework (see http://www.mitp.ru/en/predictions.html), but the time scale of the prediction uncertainty (6 months) is only of scientific interest with little or no concrete societal impact. This led many in the seismological community to conclude that earthquakes cannot be predicted (Geller, 1997; Mulargia and Geller, 2003). See however Sornette (1999) and other contributions of the 1999 debate in the journal Nature coordinated by Ian Main.

## 1.4 Solid state physics: is it the key?

Much before and well after seismology developed into a hard science, mainly based on mechanical concepts of static and dynamic deformation of materials, solid state physics may have provided another way to consider the preparatory stages to large earthquake occurrences (for instance, one can trace the connection of earthquakes and electromagnetic phenomena back to the papers of Shida, 1886, and Milne, 1890). Countless reports of precursory phenomena have been accumulated through time, first witnessing visual observations, then recorded by an ever increasing number of ground stations or remote satellites. Those precursory signals are thought to reflect the time-varying processes associated with the slow tectonic stress accumulation in the Earth's crust. Such reported pre-seismic earthquake precursors recorded on the ground or from space are compiled in Table 1 and concern very diverse categories such as variations of the magnetic field, electromagnetic fluctuations over various frequency bands, gas emanation from the ground, changes of ionospheric properties, earthquake lights, night glows, up to the controversial reports of weird behavior of domestic or wild animals.

| Satellites observations | Ground stations observations |
| --- | --- |
| Thermal Infrared (TIR) anomalies | Medium-term magnetic field variations |
| Total Electron Content (TEC) anomalies | ULF emission from within the Earth crust |
| Changes in ionospheric ion concentrations | Tree potentials and ground potentials |
| Ionospheric electric field turbulences | Soil conductivity changes |
| Atmospheric Gravity Waves (AGW) | Groundwater chemistry changes |
| CO release from the ground | Trace gas such as CO release from the ground |
| Ozone formation at ground level | Radon emanation from the ground |
| VLF attenuation due to air ionization | Positive and negative air ionization |
| Mesospheric lightning | Sub-ionospheric VLF/ELF propagation |
| Lineaments in the VIS-NIR | Nightglow amd chromatic shifts |

*Table 1:* *types of anomalous signals usually reported prior to seismic events.*

Unfortunately, the seismological community never considered such precursors as being reliable or meaningful, which led to an unfortunate disconnection, if not a definitive dialog breakage, between the two communities. The main concerns and criticisms raised within the seismological community include: (i) non-seismic precursors were not properly tested in a statistical sense, since the published cases mostly referred to single case events, and were presented without much quantification of the errors (such as false alarms and missed predictions); (ii) the reported phenomena displayed quite a wide diversity but no clear physical model had been formulated that could explain them in a coherent fashion. The latter argument is strongly reminiscent of Wegener's 1915 continental drift theory that reported so many clear and documented arguments, but was not accepted by the geophysical community until the 1960's when the fundamental process of mantle's convection was recognized. Solid-state earthquake physics thus still has to overcome its own Wegenerian bottleneck.

Based on decades of research, one of us (see Freund, 2010 for a review) has derived a credible, unifying theory for a physical mechanism that is capable of providing explanations for the multitude of reported pre-earthquake phenomena mentioned above. A clear synthesis has emerged that all pre-earthquake phenomena can be traced back to one fundamental physical process: the activation of electronic charges (electrons and positive holes) in rocks by the activation or break-up of peroxy defects during the ever-increasing tectonic stresses prior to any major seismic activity. The positive holes are unusual inasmuch as they are able to flow out of the stressed rock volume, into and through surrounding unstressed or less stressed rock, forming electric currents, traveling fast and far, and causing a wide range of secondary physical and chemical processes all along their way. These processes range from electrical ground potentials, stimulated infrared emission, and massive air ionization, to increased levels of carbon monoxide (CO) and ozone. The theoretical framework provided allows one to rationalize previous analyses of satellite and ground station data that were recorded before large earthquakes. These recordings seem to provide clear evidence that precursory signals tend to become measurable days, and sometimes weeks before the disasters.

The goal of this paper will thus be to present, in a shortened but pedagogical way, the microscopic theory of peroxy defects and its consequences at macroscopic, observable scales, as well as a critical review of observations themselves. Notice that many other theoretical models have been previously proposed (see for instance Gershenzon et al., 1989; Molchanov et al., 1995, 2001; Molchanov and Hayakawa, 1998; Sorokin et al., 2001; Pulinets et al., 2003), but will not be described nor discussed in the present paper to preserve its coherence and consistence. As pointed out above, the immense majority of observations have not been systematic in space nor time, so that even a synthesis of the related litterature, which amounts to really gigantic proportions, would amount to fall in a time and energy "black hole", and would not even be fully convincing. Such partial work compilations already exist (Hauksson, 1981; Cicerone et al., 2009) and will be cited when relevant in this paper. We shall thus provide instead a non-exhaustive review of observations in the sense that we shall focus on previously published results obtained using continuous measurements over a time period that allows one to correlate them with a significant number of events occurring within a well-defined spatial domain. The very last section will provide a discussion about these observations, as well as a presentation of our novel ideas about the way this research should continue, insisting on the proper quantitative analysis and necessary statistical testing of the correlations between non-seismic observations and earthquakes.

# II-Theoretical model: peroxy defects flow in stressed rock

We shall now present the mechanism within which we propose that most if not all reported non-seismic precursory phenomena are rooted. We choose to not present all details as they will be more meticulously described in a forthcoming book, together with many laboratory experimental results.

## 2.1 Peroxy bonds and defects in rocks

The most common minerals in the Earth's crust are silicates including quartz, feldspar, mica, amphibole, pyroxene, olivine…, abundantly present in igneous and metamorphic crustal rocks. Their structures link oxygen anions $O^{2-}$ with $Si^{4+}$ or $Al^{3+}$, typically forming $(X.Y)O_4$ or $O_3X-O-YO_3$ entities, where X and Y can be $Si^{4+}$ or $Al^{3+}$ (Freund, 1985; Freund, 2010). These minerals contain a type of defects that has largely escaped attention: peroxy links, $O_3X-OO-YO_3$, where one $O^{2-}$ is replaced by a pair of $O^-$, forming an $O^-–O^-$ bond. Special about peroxy defects is that, while they are hard to detect in their inactive state, their presence has far-reaching consequences for the physical properties of minerals and rocks, specifically for their electrical response to stress and other variables.

Peroxy defects were first observed in nominally highest purity (99.99%) melt-grown MgO single crystals [*F. Freund and Wengeler*, 1982]. They were shown to derive from the incorporation of traces of fluid phase components such as $H_2O$ into the MgO matrix during crystallization from an ever so slightly fluid-laden melt. The solute $H_2O$ turn into $OH^-$. During cooling, in the temperature range around 500°C, $OH^-$ pairs at $Mg^{2+}$ vacancy sites undergo an electronic rearrangement in the form of a redox conversion, whereby the hydroxyl protons extract one electron from their respective parent hydroxyls, turning into H, which combine to $H_2$, while the hydroxyl oxygens – now in the 1– valence state – combine to $O_2^{2-}$. This redox conversion has been confirmed by replacing $OH^-$ with $OD^-$ [*F. Freund et al.*, 1982]. Evidence for peroxy defects was subsequently obtained for silica and silicates such as feldspars, pyroxenes, olivine etc. and a variety of rocks (F. Freund and Masuda, 1991; F. Freund and Oberheuser, 1986; F.T. Freund, 2003; Lerski et al., 1988).

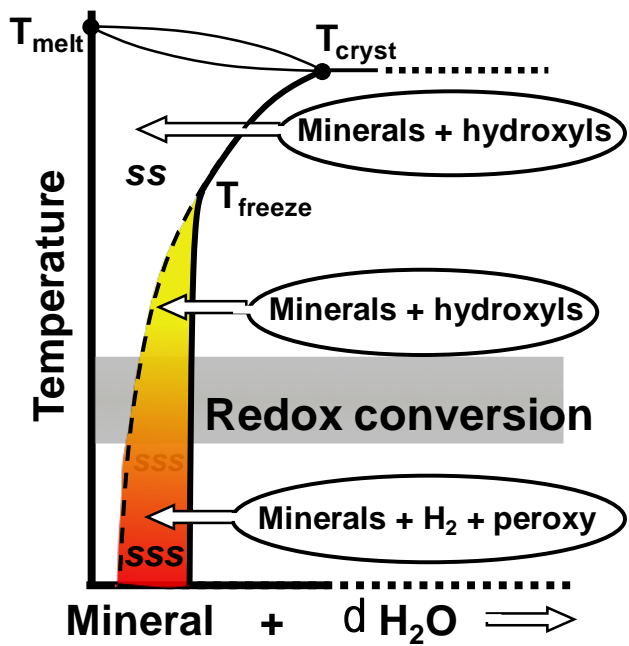

Figure 1: Part of a 2-component phase diagram "Mineral + H₂O", indicating that the melting point of the dry system $T_{melt}$ is lowered to $T_{cryst}$, the crystallization temperature, through formation of a solid solution (ss). The ss stability field shrinks with decreasing temperature. At $T_{freeze}$, the system freezes and leaves thermodynamic equilibrium. At this point, ss turns into a supersaturated solid solution, sss. In the sss field, a redox conversion takes place during further cooling, converting pairs of solute OH⁻ into peroxy plus H₂.

According to thermodynamic principles, when a solid crystallizes from a melt or, more specifically, when a mineral crystallizes from a magma that is naturally laden with dissolved gases or fluids, a finite concentration of the gas/fluid components will enter the solid matrix, forming a solid solution (ss) as depicted in **Figure 1**. The solid solution stability field is widest at $T_{cryst}$, the temperature of crystallization. With decreasing T, the solid solution stability field shrinks – a process that can only be achieved by diffusional processes, which allow the gas/fluid "impurities" to segregate, preferentially to dislocation, subgrain boundaries and grain surfaces. Eventually however, regardless of how slow the cooling rate, the diffusional processes cannot keep up. At this point the system freezes at $T_{freeze}$, and the solid solution turns into a supersaturated solid solution (sss) marked in **Figure 1** by yellow and red. Upon entering the supersaturated solid solution field, the system leaves thermodynamic equilibrium and enters the realm of metastability.

Under metastable conditions, reactions can take place that are disallowed under equilibrium conditions. Of interest here is an electronic rearrangement known as a redox conversion, which involves pairs of solute hydroxyls, OH⁻ or $O_3$(X,Y)-OH. In a silicate matrix the solute species is $O_3$Si-OH. Upon cooling to temperatures around 500°C, these hydroxyl pairs undergo a redox conversion, in the course of which each hydroxyl proton, H⁺, steals an electron from its parent hydroxyl oxygen, $O^{2-}$: OH⁻ ⇔ O⁻ + H. The two H combine to H₂, while the two O⁻ snap together to form an O⁻–O⁻ bond. In the case of MgO, this becomes a peroxy anion, $O_2^{2-}$. In the case of silicate matrices, it becomes a peroxy defect, generically $O_3$Si-OO-Si$O_3$.

$$O_3Si\text{-}OH \; HO\text{-}SiO_3 \Leftrightarrow O_3Si\text{-}OO\text{-}SiO_3 + (H_2)_i \qquad [1]$$

The redox conversion is reversible as long as the H₂ molecules remain at or close to the site, where they have been born, here marked by subscript i for "interstitial". Since interstitial H₂ is diffusively mobile, even in dense mineral matrices, they may diffuse away, here marked by an arrow, making this redox conversion irreversible:

$$O_3Si\text{-}OH \; HO\text{-}SiO_3 \Rightarrow O_3Si\text{-}OO\text{-}SiO_3 + H_2\Uparrow \qquad [2]$$

The temperature interval, in which this redox conversion takes place, around 500°C, is marked in gray in **Figure 1**.

Introduced into the mineral structures during cooling over geological times along the geotherm, peroxy defects such as $O_3Si$-OO-$SiO_3$ replacing $O_3Si$-O-$SiO_3$ can therefore be expected to exist in essentially all igneous and high-grade metamorphic rocks. When igneous rocks are transported to the Earth surface and erode, detrital mineral grains laden with peroxy defects such as quartz, feldspars etc. become incorporated into sedimentary rocks. As a consequence, even sedimentary rocks commonly contain minerals with peroxy bonds.

The presence of peroxy defects is an important geophysical factor in the sense that they will affect, even control, many features of the rocks, especially their electric transport properties. Mechanical stresses are highly effective in perturbing peroxy defects that are sitting at grain boundaries or may even bridge grain boundaries. Mechanical stresses cause grains to slide relative to each other. Any ever so slight movement will bend peroxy bonds and cause them to break up.

Electric charges in rocks are either ionic or electronic. Those that are electronic fall into two categories: electrons (usually noted as e') and positive holes or p-holes, the latter standing for a defect electron (usually noted as h$^\bullet$). Crustal rocks mostly consist of silicate minerals or contain detrital silicate minerals, characterized by $O^{2-}$ anions. However, as argued above, there will always be some $O^{2-}$ that have given away an electron and turned into $O^-$. Two $O^-$ atoms bond together in a peroxy link represent a pair of trapped p-holes, which are dormant and electrically inactive in their dormant state.

As long as peroxy bonds are intact, they are electrically inactive. When peroxy bonds are perturbed, they can break up. A type of perturbation that leads to  During the break-up an electron, e', is transferred from a neighboring $O^{2-}$ into the broken peroxy bond. In eq. [3] the peroxy bond is represented by two dots, **:**, each dot standing for a hole state. During break-up the electron becomes trapped in the now broken peroxy bond, here represented by a single dot, •:

$$O_3Si\text{-}O\mathbf{:}O\text{-}SiO_3 + O^{2-} \Leftrightarrow O_3Si\text{-}O\bullet O\text{-}SiO_3 + O^- \qquad [3]$$
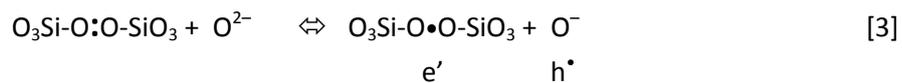

At the same time the $O^{2-}$ that has donated the electron turns into an $O^-$. This quasi-free $O^-$ represents a defect electron in the oxygen anion sublattice, e.g. a hole-type charge carrier, symbolized by h$^\bullet$, which we call a "positive hole" or p-hole for short.

The transport properties of the trapped electron e' and of the quasi-free hole, h$^{\bullet\prime}$, are of interest In the constext of earthquake-related electrical properties.

The energy levels of the unbroken peroxy defects are associated with O 2sp-symmetry states that form the upper edge of the valence band. Likewise, the new e' and h$^\bullet$ states, created during the break-up of the peroxy bonds, exist at or near the upper edge of the valence band. The e' become trapped by shifting downward to new energy levels slightly below the surface of the valence band. Their conjugated mirror states e'* shift upward into the band gap, slightly above the edge of the valence band. It is important to note that the e' and e'* states are available only where peroxy bond breakage occurs, i.e. where mechanical stresses are applied the rocks, which cause peroxy bonds to break. Hence, the e' are mobile only within the stressed rock volume.

By contrast, the h$^\bullet$ are associated with energy levels that form the very edge of the valence band. Their wavefunctions are highly delocalized and they have the remarkable ability to spread out of the stressed rock volume.

Subjecting a rock to deviatoric stress reveals the existence of those dormant p-holes (Freund et al., 2006). If the stressing rate is sufficiently small, which is the case in tectonics, rock will at first deform continuously with a visco-elasto-plastic bulk rheology (i.e. in the ductile regime, as opposed to the brittle regime where deformation is mainly accommodated by elasticity and the formation of cracks and fractures). Most of the ductile deformation is irreversible and consists, at the microscopic level, in sliding of grains relative to each other and in motion of dislocations, a thermally activated process. To the extent peroxy bonds decorate grain boundaries or dislocations, every so slight motion of mineral grains relative to each other will tend to break the $O^-$-$O^-$ bonds. As described by eq [3] the broken peroxy bonds will take over an electron from a neighboring site, for instance a nearby $O^{2-}$ acting as donor. While the electron becomes trapped in the broken peroxy bond, the donor $O^{2-}$ turns into an $O^-$, equivalent to a defect electron on the $O^{2-}$ sublattice, i.e. a p-hole. This p-hole state is not bound to the broken peroxy bond but can diffuse away via a phonon-coupled electron hopping mechanism [*Shluger et al.*, 1992] at a speed on the order of 100 m.s$^{-1}$ [*John Scoville et al.*, 2015]. The p-holes are capable of traveling over large distances. Since their speed of propagation is controlled by diffusion, they will slow down, if they propagate in a plane or into 4π space [*John Scoville et al.*, 2014; *John Scoville et al.*, 2015].

The break-up of peroxy bonds will naturally enhance the electric conductivity of the rocks, in particular within the stressed rock volume, where e' and h$^\bullet$ jointly act as mobile charge carriers, less so in the surrounding rocks, through which only h$^\bullet$ can travel. To the extent that the presence of the e' and h$^\bullet$ charge carriers affect other physical properties of rocks, for instance the speed of P and S waves, this process may provide a new and different explanation of the widely reported pre-earthquake changes in the $V_P/V_S$ ratios (Catchings, 1999; Scafidia et al., 2009). The most common explanations considered so far are that, when rocks deep in the crust are subjected to stress, they undergo microfracturing,. This process is thought to allow fluids to penetrate, which would lead to changes in the speed with which P and S waves propagate (Kranz, 1983; Scholz, 1968). The alternative explanation offered here is that, when peroxy bonds become activated and p-hole charge carriers with highly delocalized wavefunctions become activated, the average bonds between oxygen anions and cations become slightly less ionic. Increased covalency makes the bonds slightly more ductile, leading to changes in the mechanical properties of the rocks that have been described as "softening" (Freund et al., 2010).

Once e' and h$^\bullet$ charge carriers become activated in a given rock volume, they will start to recombine, returning to the inactive peroxy state. By conducting laboratory experiments at different stress rates spanning 8 orders of magnitude, it has been shown that, inside the stressed rock volume, the lifetimes of stress-activated p-holes vary widely from milliseconds to several months [*John Scoville et al.*, 2015]. If the stress rates are very high, the number of p-hole charge carriers activated and available to flow out of the stressed subvolume is very high. Once outside the stressed subvolume, the p-holes are expected to have indeterminate lifetimes, possibly very long, allowing them to propagate far afield. This suggests that observable signals rooted in the physics of p-hole migration may display a wide range of patterns at different stress rates.

In laboratory experiments, uniform loading of a sample of dry granite under uniaxial stress conditions leads first to a nearly 5-fold increase of its electric conductivity until the stress reaches a moderate value of about 25 MPa. Beyond this stress value, the conductivity increases much more gradually and indeed quickly reaches an almost constant value (Freund, 2010). In another experiment, only a small part of a rock slab is loaded while the rest is kept unstressed (Freund, 2010). Voltage is measured across the plate at two different locations (within stressed and unstressed domains). The stressed domain shows a monotonic but irregular increase of the measured voltage, up to a very sharp peak whose onset and maximum value coincide with major cracking events preceding the final failure of the sample. In the unstressed domain, the temporal pattern of voltage is completely different: it first rises very quickly to its maximum value as soon as loading is applied on the other part of the sample, then decreases irregularly with large fluctuations to finally reach its pre-stress background value, and reverses slightly its sign just before failure. Freund (2010) showed that the most significant part of the electric signal is thus observed before reaching failure stress, suggesting that electrical anomalies in nature would not necessarily continue to build up towards earthquakes but precede them by a finite time depending on the loading rate. In a similar but constant stress experiment, Freund et al. (2006) observed that the current intensity measured across the sample mimicked closely the time variation of the imposed stress.

As already mentioned, the existence and mobility of p-holes should increase the electric conductivity. In a heating experiments (Kathrein and Freund, 1983; Freund, 2010) conducted with MgO single crystals, the number of p-holes started to increase around 430° and further increase over the 450-600°C temperature range, causing the conductivity of the MgO single crystals to increase by up to 6 orders of magnitude. The current carried by these p-holes has a characteristic 1eV activation energy. Interpreting laboratory experiments by Parkhomenko and Bondarenko (1986) who monitored the electric conductivity of various rock samples as a function of temperature, Freund (2010) notices a similar energy activation from 200 to 600°C, suggesting that the same migration of p-holes process also holds in crustal rocks. Transposing this temperature range onto the average geotherm, i.e. the depth–temperature profile, in the stable continental crust, this roughly corresponds to the depth range of 7-40 km. Incidentally, this fits well with the depth range over which most earthquakes are observed to nucleate.

### 2.2 Flow of positive holes in the crust

Freund (2002) and Freund et al. (2006) suggested that, when activated by stress (and temperature, but we shall from now on only focus on stress), nucleated p-holes start to flow towards the less stressed regions while electrons will remain trapped locally. The unstressed rock volume thus becomes positively charged relative to the stressed volume, making the system behave like a battery. The potential difference that emerges creates an electric field counteracting the flow of p-holes. As the latter naturally repel each other, positive charges will pile up at the surface of the Earth. A side-effect is a significant electrical conductivity increase within the stressed volume, and a smaller increase across the unstressed domain (Freund, 2010).

No sustained outflow of p-holes can occur when the battery circuit is not closed to allow for a return current. In laboratory experiments it is easy to achieve circuit closure by connecting the stressed and unstressed parts of a given rock sample with a wire. Electrons in the stressed portion of the rock then use this wire to flow to the unstressed portion and recombine with the p-holes that have traversed the unstressed rock. In the field the situation is more complicated. Three scenarios can be envisioned.

First, in the case of large to very large earthquakes the actively stressed rock volume may extend downward to the deeper layers of the crust, where the temperatures are sufficiently high so that electrons are thermally activated causing the rocks to become electron-conductive. In this case an electron current can flow deep in the crust paralleling the stress-activated p-hole current and thereby closing the circuit (Freund, 2007a, 2007b).

Second, if an electrolytically conductive path exists, for instance through the water-saturated gouge along a fault plane, circuit closure can be achieved through the flow of $H_3O^+$ and water-soluble ions (Freund, 2009). This mechanism is plausible as active faults may permit deep penetration of water along fault planes.

Third, if large-scale air ionization occur at the Earth surface, the conductivity of the air may become large enough to provide for a return path for the p–hole charges flowing in the crust.

## 2.3 Accumulation of p-holes at the Earth surface and air ionization

While accumulating at the free surface the positive charges form thin surface/subsurface charge layers associated with steep electric (E) fields. Due to the intrinsic nature of E fields (represented mathematically as Laplacian fields), corners, edges, or any other positive (i.e. upward) topographic fluctuation (such as hills or high mountains in areas of tectonic convergence) amplify these E fields, which can locally reach values as high as a few millions $V.cm^{-1}$.

As these high E fields build up, they will lead to two processes (Freund et al., 2009):

(i) Field ionization of air molecules. Among the main constituents of air $O_2$ has the lowest ionization potential and, hence, is expected to become most easily field-ionized to $O_2^+$.
(ii) Corona discharges. When the local E fields become so large that they can accelerate free electrons (which are always present due to cosmic rays and radioactive decay processes) to energies sufficiently high to impact-ionize gas neutrals, discharge avalanches will occur, commonly known as corona discharges.

Air ionization has been shown to set in once the surface potential reaches +3V (Freund et al., 2009). Further influx of positive holes leads to fluctuations of the surface potential, indicative of impulsive field-ionization of positive airborne ions at the surface and injection of electrons into the surface. If the influx continues, corona discharges are triggered, leading to a break-down of the positive surface potential (Freund et al., 2009).

It is expected that topographic fluctuations where ionization is maximum will then allow a lot of air electrons to be driven to the rock surface from above, causing the upward flow of p-holes from the bulk rock to increase correspondingly. Laboratory experiments show that the generation of positive airborne ions at the surface of stressed gabbro samples displays a pulse-like dynamics and also a major peak at sample failure, due to the sudden release of a large amount of airborne ions by the fracture surface. The reversal of the surface potential before failure is accompanied by light flashes emanating from the rock sample as well as by radio-frequency pulses, confirming the occurrence of corona discharges so that conditions are fulfilled to get air ionization.

This ultimate step in the life of p-holes is evidenced by another set of laboratory observations. Close to the free surface p-holes can recombine and return to the peroxy state. The recombination is an exothermal process, in the course of which part of the energy is recovered that had been expended in the stressed rock to break the peroxy bonds. This energy leads to vibrationally highly excited states of the newly formed peroxy bonds, which de-excite by emitting photons at discrete energies corresponding to the downward transitions in the vibrational manifold (Ricci et al., 2001). an experimental confirmation of these infrared "hot bands" has been reported for anorthosite (Freund, 2007).

## 2.4 First-order predictions of the model

The model presented above provides a very plausible overview of the 'life, works, and death' of positive charge carriers in the Earth crust, namely the p-holes. Unfortunately, there exists no direct means of investigation in order to validate this scenario, and microscopic entities such as p-holes are not detectable *in situ*. However, if rock material in geological conditions obeys the physical picture described in the previous section when being slowly loaded by tectonic stress, the resulting positive charges accumulation and massive air ionization at the free surface would induce a wealth of corollary phenomena that we shall now describe in more details. Most of them are fairly measurable and have been claimed to be widely observed before seismic events of various magnitudes. Published works reporting them will be reviewed in the next section.

The most obvious expected consequence of the arrival of p-holes at the free surface would be an increase of the electric conductivity of the most superficial soil layers, i.e. a change of a physical properties. A more subtle consequence is of chemical nature, as soils generally feature high contents of organic matter, i.e. of carbon atoms. The latter is often present in the form of triple bonded carbon, which is able to retain Radon ($^{222}$Rn) atoms. Radon is a noble gas with a half-life time of about 3.8 days, generated from the radioactive decay of Radium ($^{226}$Ra), and which can display chemical reactivity (Li et al., 2008). As Radium itself is a decay product of Uranium, it follows that Radon is geologically mainly confined over continental areas. Freund (2010) speculates that, because p-holes are highly oxidizing, they should oxidize triple-bonded carbon to double-bonded carbon. As the latter is unable to retain Radon atoms within the soil, this allows the release of Radon atoms, which can then freely percolate within the soil and escape at the free surface. Such Radon emanation in the vicinity of future epicenters have been reported a countless number of times as a genuine earthquake precursor. Note that observations of Radon concentration increase certainly reflect very local conditions as its short life-time doesn't allow it to travel over very large distances, certainly a few meters at most (Woith, 2015). Another expression of the highly oxidizing nature of p-holes can be found in the fact that, prior to major earthquakes, carbon monoxide, CO, has been found to emanate from the ground, probably due to the oxidative interaction of p-holes with organic matter in the soil [*Singh et al.*, 2010].

As observed in some of the laboratory experiments reported in the previous section, infrared emissions are expected to occur when p-holes recombine with electrons at the free surface. As such recombination should occur at higher rates at narrow topographic heights, where the inward flow of p-holes is predicted to be the largest, we should thus observe a correlation between the topography and Infrared emissions, which are usually misinterpreted as an increase in actual ground temperature. Such a model rationalizes the night-time thermal anomalies that have been observed using Infrared satellite imagery (Ouzounov et al., 2006; Ouzounov and Freund, 2004; Qiang et al., 1999; Saraf et al., 2008; Singh, 2008; Tramutoli et al., 2005; Tronin, 2000; Tronin et al., 2004), without the need to use a specifically thermal model (examples of which can be found in Pulinets et al., 2006; Saraf et al., 2008; Singh, 2008; Tramutoli et al., 2005; Tronin, 1999). An excellent example of strong pre-earthquake thermal infrared emission from the mountain tops has recently been reported for the case of the magnitude 6.3 L'Aquila earthquake [*Luca Piroddi et al.*, 2014a; *L. Piroddi et al.*, 2014b]. Though pre-earthquake radon emanation might be sufficient to locally change the heat capacity of the air, this does not the fact that the most intense infrared emissions come from topographic highs and not from the valleys that are dissected by active faults.

Increasing the electric field at the ground surface up to the possible triggering of corona discharges suggests that transient phenomena usually reported as earthquake lights may also occur (Galli, 1910; Losseva and Nemchinov, 2005; Mack, 1912; St Laurent, 2000; Terada, 1931; Tsukuda, 1997; Derr, 1986). Such corona discharge are also speculated to generate a significant amount of RF (Radio Frequency) noise (Freund, 2010) which should be recorded. A more speculative consequence is the water droplet condensation expected to occur around nuclei constituted by airborne ions. This condensation will be accompanied by a release of latent heat, causing the rise of this air mass (Dunajecka and Pulinets, 2005). Under favorable conditions of humidity, clouds can form and remain close to the future epicenter zone. Such cloud formation has been documented before earthquakes (Lu, 1988; Ondoh, 2003; Tramutoli, 1998; Guo and Wang, 2008).

At last, once the massive ionization of air occurs, this would lead to an upward migration of charged particles, i.e. to a vertical current flow in the atmosphere that Freund (2010) estimates to be of the order of 10-100 A km$^{-2}$ and which would produce noticeable electromagnetic anomalies. Those anomalies would not be restricted to the atmosphere as the ascending positive charges would then also pull downward electrons located at the bottom of the ionosphere, thus modify its physical properties by influencing the vertical distribution of electrons and ions in the ionospheric plasma (Chen et al., 1999; Hayakawa, 2007; Hayakawa et al., 2005; Liperovsky et al., 2000; Pulinets, 2007; Sorokin et al., 2006; Zakharenkova et al., 2007).

## III-Empirical tests

### 3.1 Observations

The previous section exposed a consistent theoretical model of stress-dependent electric charge activation and migration, potentially leading to a wealth of phenomena which would be observable prior to earthquakes. We are very aware that the upscaling from the microscopic and laboratory scales up to the size of interest for natural earthquakes is far from obvious, in view of the structural complexity of the Earth's crust. The latter is crisscrossed by innumerable cracks, fractures, joints and faults over the full observable range of scales (typically from microns up to thousands of kilometers). Those discontinuities are thought to be often filled by fluids of various chemical compositions. This pervasive disorder might itself induce a high complexity in the geometry of the path over which electric charges may travel in rock. Charge flow might be highly focused in some zones and nearly completely screened in others, possibly leading to a very heterogeneous structure of the distribution of charges close to the surface. We are still far from a complete forward modelling of such a process, and we should be aware that observations can by no means be as smooth in space, time or amplitude as those that could be deduced from a similar process in a homogeneous medium. The association of anomalous phenomena and earthquakes will thus necessarily be imperfect, which is why we chose to report here only works dealing with systematic analyses that allow one to assess the statistical significance of the underlying physical assumptions.

To be fully consistent with the previous section where we explained the theoretical model, we shall review only some of the various phenomena that can be directly predicted to hold, namely: radon gas emanations, corona discharges, thermal infrared emissions, air ionization, ion and electron content in the ionosphere, and electro-magnetic anomalies.

One should keep in mind that all these observations rely on different methods of measurement, which can make interpretation difficult. For instance, ground stations are usually run continuously in time, but the spatially covered area is ill-defined. Several such stations are generally run simultaneously (defining a local network), but the spatial area they cover is similarly blurred, as some recorded anomalies might have their source located far outside of the network. Correlations with seismicity are thus difficult to assess. On the other hand, satellites have the advantage to perform repetitive recordings over much wider areas on the Earth over long time periods. Some satellites are stationary or define a constellation so that almost any point on Earth can have a measurement of a given parameter continuously. This is for example the case for GPS data, which can be used to compute the Total Electron Content. Some other satellites are single and non-stationary, such as DEMETER. In that case, the embarked instruments do not provide a continuous recording of a given parameter at all locations, but a continuous sampling along the satellite's trajectory. It follows that the sampling above the location of a given point on Earth will turn out to be highly discontinuous in time.

Most of the papers dealing with non-seismic precursors refer to the work of Dobrovolsky et al. (1979) in order to check the consistency between the size of an earthquake and the maximum distance up to which anomalies have been reported. Based on the compilation of previously published data, this paper proposes that this maximum distance between an earthquake source and a precursor is given by $D=10^{0.43M}$ in km. Dobrovolsky et al. (1979) proposed a theoretical explanation to this empirical "law" (which is certainly another abusive and unfortunate terminology), by assuming that the earthquake preparation zone scales with the size of the upcoming event, and can be modelled as a soft inclusion in an elastic medium that perturbs the distribution of stresses and strains. They show that the aforementioned precursory distance corresponds to a strain perturbation of about $10^{-8}$. The idea behind the soft inclusion model is that a multitude of microcracks nucleate or open close to the future event, so that the mechanism underlying the various precursors should be of mechanical nature. Similarly, in the time domain, Rikitake (1986) proposed a relationship between the earthquake magnitude and its precursor time T: $\log(T)=0.76M-1.83$, based on the observation of many reported precursors of various nature.

As a final remark before starting our review of the main non-seismic precursors, we stress that only papers written in english have been considered, whereas most of those anomalies have been studied for decades in China, India, Japan, Russia, Taiwan, etc. so that, many papers and reports published in the corresponding languages will not be summarized or listed, representing an unfortunate western bias.

### 3.3 Radon gas emanations

The measurement of Radon gas content is one of the most often reported earthquake precursors, and is generally based on the detection of the alpha particles that are emitted by the radon decay. The very first measurements have been performed by Shiratoi (1927) and Imamura (1947) in ground water, and by Hatuda (1953) in soil along a Japanese active fault. Radon concentrations have often been reported to increase (sometimes up to a factor 10) before seismic events, on the time scale of days to months, either over large areas or in the close surroundings of an active fault (Okabe, 1956; Chyi et al., 2002; Inan et al., 2008; King, 1980; Nagarajaa et al., 2003; Tsvetkova et al., 2001; Yasuoka et al., 2009). Riggio and Santulin (2015) classify Radon anomalies into two categories according to their shapes. Type A anomalies correspond to a slow but regular drift that can take place over several years. Type B corresponds to shorter-lived anomalies (with a duration of few hours to days) that can precede earthquakes. The latter anomalies are reported to be either positive or negative. An important challenge is the removal of external influences such as meteorological conditions, as Woith (2015) claims that such externally induced anomalies look strikingly similar to those associated with seismotectonic processes.

Radon indeed seems to be a very sensitive *in situ* stress gauge. For instance, Teng and McElrath (1977) report an experiment where Radon concentration in a hot spring is measured every 2 hours over a total time period of 9 days. A simple harmonic analysis suggests that the dataset features diurnal as well as semidiurnal fluctuations, i.e. closely related to Earth tides. Shapiro et al. (1980) analyze 20 months of continuous monitoring at the Kresge site in Pasadena, covering 1977 and 1978. Data are sampled three times a day, allowing them to show evidence of an annual cycle, which they interpret as being due to the thermoelastic stresses emerging from changes in the subsurface temperature. The corresponding estimated strain is of the order of $5 \times 10^{-6}$, i.e. within the range proposed by Dobrovolsky et al. (1979). Trique et al. (1999) study two lakes in the French Alps, with water levels varying with amplitudes of, respectively, 50m and 70m. The two sites are equipped with instruments allowing to measure Radon emanations, electric potential variations, as well as the strain induced by the fluctuating water levels over a period of nearly 3 years. They find that Radon emanation bursts are highly correlated with episodes of strain acceleration: positive peaks of both time series occur within 10 days of each other in 63% of the cases (16 events in all; the score decreases to 27% if one distributes the same number of radon bursts randomly in time). A similar but slightly weaker association is found for fluctuations of the electric potential (the correlation score is 53%, and decreases down to 30% when randomizing the electric data), while the tiltmeters indicate strain amplitudes of about 5 to $7 \times 10^{-6}$.

A rough correlation of Radon emanation and seismicity is documented by Inan et al. (2008), who report a continuous monitoring at a soil radon station along the North Anatolian fault during the full 2002 year. They notice that many peaks of concentration are correlated with the occurrence of M≥4 events within a radius of 100km from the recording station. Interestingly, all earthquakes occur between January and October 2002, when the radon signal is characterized by quite large fluctuations. From early October to the end of December, the radon signal returns back to a low and nearly constant background level, while seismicity coincidentally shuts down. Some seismic events occur right at the same time as the Radon peaks, sometimes a few days afterwards, or shortly before such peaks. One should also mention that two of the nine detected seismic events clearly occur at a local minimum of the Radon signal.

Instead of considering cases of continuous monitoring and systematic correlation with earthquakes, some authors prefer to perform statistical analyses after building a compilation of previously published studies. In most of the cases, anomaly detection is performed using the method of Igarashi and Wakita (1990), which consists in estimating the long term average of the signal, and looking for times when the signal deviates by more than two standard deviations. Hauksson (1981) provides a compilation of many data collected in the literature, devoted to single case analyses, in order to put in evidence general relationships between the properties of the recorded anomalies and the magnitude of the associated earthquakes. As he points out, in most analyses that have been performed by various teams across the world, the rate of false alarms as well as the absence of anomalies are generally not reported, which certainly biases his statistics. Such a data compilation is thus by no means equivalent to a systematic analysis. Hauksson (1981) claims that, in the case of Radon anomalies, the relationship of Rikitake (1976) for the anomaly lead time is much more often the exception than the rule. His compilation suggests that the amplitude of the Radon anomaly for events with magnitude between 6.0 and 8.0 tends to peak at distances of 200 to 500km from the epicenter, but does not give any indication about its azimuthal position. We shall see below that such a distance also emerges in the case of electric field anomalies recorded by satellites. This distance is observed to grow with time and with the magnitude of the event, which is compatible with the empirical law of Dobrovolsky et al. (1979), as the zone where Radon anomalies exist corresponds to strain values of about $10^{-8}$ to $10^{-6}$. However, the amplitudes of the anomalies do not show any correlation with the magnitude of the event. Woith (2015) also provides such a meta-analysis of 93 papers relating radon precursor anomalies, and suggests that the precursor time interval before events can reach up to 18 years, but data selection suffers the same limitations as Hauksson (1981). Cicerone et al. (2009) provide a statistical analysis of 125 previously published observations linked to 86 earthquakes. They conclude that the temporal organization of the anomalies does not allow to foretell the time of the event, despite the fact that most anomalies seem to occur within the previous month and last less than 200 days. Larger precursory times (with maximum values varying from 0 to 200 days when magnitude increases from 2 to 8) as well as longer duration of the Radon anomalies (with corresponding maximum values from 200 to 1000 days) appear to be correlated with larger magnitude seismic events. They also suggest that larger anomalies tend to occur closer to the epicenter, and confirm that the amplitudes do not correlate with the magnitude of the event. Data are anyway very scattered, and most of the anomalies exhibit amplitudes of the order of 50% to 100% of the background level. See also Petraki et al. (2015) for another similar review of published works.

Looking at a continuous recording, Teng (1980) provides a contrasting account of the use of Radon anomalies for earthquake prediction purposes. He first analyses data recorded along the locked part of the San Andreas fault, where Radon is sampled weekly since 1974. At the time of his paper, he notices that no clear correlation can be assessed between earthquakes and the Radon signal. For instance, at the Switzer Camp station, three positive anomalies are detected, but only one seems to correspond to a seismic event. Moreover, no correlation among anomalies recorded at different stations seems to exist. He also reports the case of the Kutzan station, located West of the Szechuan province in China, close to a large active fault where nine events with magnitude between 5.2 and 7.9 occurred during the time period 1972-1976. Prior to eight of them, anomalous spikes in Radon concentration have been recorded from 6 to 13 days before the events, and only one anomaly was not followed by any seismic event. Those anomalies stood from 36% to 120% above the long-term average level. Teng (1980) acknowledges that many anomalies worldwide do not seem to be linked to any event, so that the rate of false alarms in a prediction set-up might be quite high.

Shapiro et al. (1980), whose experiment has been described above, show that only 3 out of a set of 11 events with M≥2.0 and depth within 2-15 km, within a 25km distance from the station, have been preceded by a Radon anomaly. Four are reported to be preceded by possible fluctuations due to some external cause (such as rain), while four came without any precursory signal. Note that, in this work, anomalies have been detected by a simple visual inspection.

Hauksson and Goddard (1981) collect data from nine stations located in Iceland, with most stations being spaced by less than 15km, sampled once a week in 1978 and 1979. In order to associate recorded Radon anomalies and observed earthquakes, they propose a relationship to relate the minimum earthquake magnitude M that is able to trigger an anomaly at a recording station, as a function of the epicenter-to-station distance D. They find that M≥2.4 $\log_{10}(D)$-0.43 (which, when inverted, yields an estimation of D that is a bit larger than the one initially proposed by Dobrovolsky et al., 1979). This relationship is fitted on data reported in the literature for large earthquakes occurring in China, USSR and Japan, and simply extrapolated to smaller magnitudes for their observations in Iceland. When earthquakes are clustered, only the largest one is taken into account, or the first one of the sequence if their magnitudes are similar. Radon anomalies are then detected using the criterion of Igarashi and Wakita (1990). When considering all events (23 earthquakes with magnitude ≥1.0) and stations, they are then left with a database of 57 potential observable radon anomalies, from which they deduce a set of 9 observed precursory anomalies, 48 cases of lack of anomaly (false negatives), and 7 false alarms (or false positives). They also observe that the amplitude of the anomaly does not seem to change with distance. Anomalies last from about 2 weeks to a month, their amplitudes relative to the background level being about 40% to 380%. When considering only events with M≥2, 65% of the latter can be associated with at least one anomaly, and both the duration of anomalies and the maximum distance where they are observed increase with the size of the event.

Steinitz et al. (2003) report an eight years (1994-2002) experiment near the Dead Sea rift fault, where 796 events with magnitude between 0 and 4.6 have been detected. They first remove events with a location uncertainty larger than 4km (thus selecting 82% of the whole catalog) and exclude events that appear to be clustered with previous ones. They compute a running average of the Radon time series using a 25h long window. From this time series, they determine the local minima and maxima. For each locally maximum value, they compute the ratio between that value and the preceding minimum value (which is considered as the onset of the corresponding anomaly). They then show that only anomalies ≥1.9 are correlated with earthquakes, which constitutes a set of 110 anomalies, considering events occurring as far as 270km from the monitoring station. They then show that events occurring outside of the Dead Sea rift valley do not show any specific clustering relative to the anomalies, whereas those occurring within the rift do display such a clustering within the three days following the onset of the anomaly. This result is validated by generating random earthquake sequences, showing that the natural observations have a probability of being due to chance of only p=0.006%. This result is stable with respect to the specific rules employed to remove clustered events.

Torkar et al. (2010) use the data of a Radon station in Slovenia, operating from June 2000 to January 2002, with an hourly sampling rate. For each earthquake, they compute the radius predicted by Dobrovolsky et al. (1979) and look for prior anomalies before this event within twice this distance. The time series of Radon concentration and other meteorological data during non-seismic periods are then fed into a multilayer perceptron (i.e. a multilayer neural network) in order to decipher the effect of external non-seismic environmental parameters on the Radon signal. The parameters of the neural network are then applied to the set of environmental parameters during seismic periods in order to predict what the Radon concentration should be, and compare it to the observed one. The difference between the two defines the potential anomaly. This technique then allows one to associate 10 of the seismic events (i.e. 77% of the total) to a Radon anomaly, within +/-7 days.

Let us also mention the report by Igarashi et al. (1995) of the fourfold increased Radon concentration in ground water over several months before the 1995 southern Hyogo Prefecture (Kobe) earthquake on 17 January 1995. On 8 January, 9 days before the earthquake, the radon concentration reached a peak of more than 10 times that at the beginning of the observation, before starting to decrease. Johansen et al. (1996) found that the dynamical evolution of the Radon concentration is well-represented by log-periodic accelerated peaks, suggesting a kind of critical behavior (Freund and Sornette, 2007).

### 3.4 Ground measurement of electromagnetic fields

#### *3.4.1 Electric field*
In the DC to ULF range (i.e. up to 10Hz), Myachkin et al. (1972) report variations of the electric field amplitude of the order of 100-300 mV.km$^{-1}$, during the 3-16 days before events in Kamchatka, but not systematically. Sobolev (1975) similarly noticed a decrease of the electric field prior to events in Kamchatka, using hourly means. In contrast, Miyakoshi (1985) reports an increase of amplitude on just one of the two components in Japan. This corresponds to variations at periods of a few hours to days. At higher frequencies, no such variations are noticed (Honkura et al. (1976) report daily variations of 50% at periods 60-7200s without any associated seismicity; variations of up to 100% are observed by Sims and Bostick (1969) in Texas without any event too). In Greece, Varotsos and Alexopoulos (1984a, 1984b, 1987) and Varotsos and Lazaridou (1991) reported square pulses of up to 250 mV.km$^{-1}$ preceding events by 7-260 hours (so-called Seismic Electric Signals, SES). Scaling of the observed signals on dipole gauges of different sizes are used to discriminate true signals from noise.

#### *3.4.2 Magnetic field*
The release of positive holes from a source volume generates currents that are accompanied by corresponding changes in the magnetic field. If the magnetic field changes rapidly, transient magnetic pulses are observed. In fact, it is this phenomenon that allowed Shockley et al. (1949) to verify the diffusive behavior of charges in semiconductors. Such pulses were computationally modeled in the context of positive hole flows by Scoville et al. (2015) and were found to have the characteristic diffusive form of pulses observed before several earthquakes in Peru, within an order of magnitude in amplitude and duration.

Similar pulses were observed with increased frequency prior to the M = 5.4 Alum Rock California earthquake of 30 October 2007. A magnetometer located about 2 km from the epicenter recorded a series of unipolar magnetic pulses reaching amplitudes up to 30 nT, as reported in Bleier et al. (2009). Bleier et al. (2009) shows that, in the three weeks preceding the Alum rock earthquake, the incidence of pulses was higher than any other 3-week period from 2006-2007 and that the pulse count falls rapidly after the earthquake. Bleier et al. (2009) also notes that there were no nearby lightning strikes at corresponding times, the pulses were much stronger than PC3 and PC4 geomagnetic pulsations, and were localized near the epicenter, which would not be the case for geomagnetic activity.

Johnston et al. (1973) offer a review of reported variations of the magnetic field amplitude before earthquakes, and show that their amplitudes decrease drastically with time after 1960. This is interpreted as resulting from higher quality instruments, and from the removal of anomalies due to ionospheric and magnetospheric disturbances. Honkura et al. (1976) show that, below 0.1Hz, the spectrum of the natural field has a 1/f spectrum, which explains the existence of many fluctuations superimposed on the DC component, so that anomaly detection might be difficult in this frequency range. For instance, Dea et al. (1993) successfully correlate ULF signals with 29 events with M>3.5 in California and Nevada during 18 months. However, they report associated signals for only 7 of the 67 events with M>3.5 in Southern California during the same period.

Chen et al. (2004) study the variations of the total geomagnetic field recorded by eight ground stations in Taiwan during 1989-2001 with a sampling rate of one measurement every 10 minutes (except one station, with a sampling interval of 5 minutes). They compute the yearly drift of the amplitude of the magnetic field, and data are compared with the international geomagnetic reference field model (Barton, 1977). The drift is found to be less than 5nT/yr for all stations but one after 1997, thus defining a zero isoporic zone (ZIZ) as defined by Zeng et al. (2001) who claim that 80% of events with magnitude larger than 6.0 occur within 9 months to 2.5yrs after the onset of a ZIZ. Chen et al. (2004) claim that such a ZIZ appears in Taiwan after 1997, coinciding with the onset of seismicity, while the ZIZ disappears after the Chi-Chi earthquakes and its aftershocks sequence. Zeng et al. (2002) claim that a ZIZ appeared within 2.5 years before the Haicheng and Tangshan earthquakes in China, while Ispir et al. (1976) report a similar process about 1 year before M>=6.0 events in Turkey during 1966-1969. Tazima et al. (1976) report the same behavior within 2.5 years before 80% of events in Japan during 1954-1966.

Johnston (1989) reports a continuous experiment of magnetic field monitoring along 800km of the San Andreas fault system, with only one event with M=5.2 preceded by a magnetic anomaly on two independent instruments (which happened to be the closest to the epicenter), while a nearby event with M=5.9 gave no precursory signal. The recording system of the USGS operated for 14 years.

Smith et al. (1978) failed to correlate creep events on the San Andreas fault with signals recorded by nearby magnetometers using 3 years of data. However, Johnston (1989) suggests that, when detrended, changes in creep rates over time scales of several months appear to be correlated to similar changes in the magnetic field, which is interpreted as the effect of the tectonic load. Creep events are also not associated with local variations of the electric field.

Han et al. (2014) use geomagnetic data in the ULF range (i.e. around 1Hz) recorded by a single ground station in Japan during the time period 2001-2010. The quantity they study at a given station is not a standard one and is derived from Hattori et al. (2006). They first consider each event and its associated co-seismic energy release. The quantity that such an event transfers to a station is its energy divided by the squared hypocentral distance. Coarse-graining at the scale of 1 day, they thus compute a daily cumulative energy index at each station, and consider only cumulated values larger than $10^8$ (unfortunately, no unit is provided), which are then labeled as earthquake events, and consider only seismic events with a depth shallower than 60km. They consider two spatial subgroups of events: the closest 50 ones, and the next closest 50 ones. For the magnetic signal, they only consider data recorded from 2:30am to 4:00am in order to remove external magnetic pollution. The recording of the vertical component sampled at 1Hz is then wavelet-transformed at about 0.01Hz, and its power is coarse-grained at a daily scale too. Another station, located in an aseismic area, is used as a reference in order to remove the effect of global magnetic fluctuations. Using a simple linear regression model to link both magnetic signals, they are able to compute a reference signal at the observation station, and simply use the ratio between the observed and modeled values at that station in order to define an anomaly. An anomaly is defined when the ratio is beyond the median of its distribution plus 1.5 times its interquantile range. This thus means that only energy enhancements are considered as anomalous, without any consideration of energy drops. They are thus left with a set of 324 anomalies. They then use a superposed epoch analysis by considering consecutively each earthquake as the origin of time, in order to check the average distribution of anomalous geomagnetic days before and after the event. A similar procedure is conducted by randomizing the times of earthquakes using a uniform distribution. Unfortunately, they do not decluster the events (or only partially, thanks to their daily coarsegraining). Repeating this last step several times allows them to compute error bars on the surrogate sets. The results show a clear clustering of anomalies in the two weeks before and 2 days after earthquake events. Coarse graining at scale of 5 days yields an anomalous period of 6-15 days before earthquake events. When considering the dataset that is the most distant to the station, those correlations disappear, suggesting that only events close to a station can generate magnetic anomalies. Varying the threshold of the energy index, they find that only earthquake events which induce values larger than $10^7$ are significantly associated with anomalies that precede them. They propose to quantify their results by plotting them onto a Molchan error diagram, considering only the case where, once an anomaly is evidenced, an alarm is triggered within a time window extending 11-15 days after it. This should lead to a single point on the diagram, but the authors plot a continuous line from which they deduce that their gain is around 1.6 when compared to a random prediction. The parameter used to get that curve is not mentioned, so their conclusion is a bit unclear. We can guess that the parameter is the threshold value defining if an anomaly is observed.

In a subsequent paper, Han et al. (2017) use exactly the same dataset and processing in order to test for the influence of the leading time and window size of the alarms to get the best prediction when an anomaly is observed. They use a slightly different version of the Molchan diagram by plotting the ratio of the correctly predicted events versus the ratio of the alarm time rate. As a bonus, they compute the 90% and 95% confidence level curves of the random guess case in order to check the significance of a prediction that is observed to be better than random. They then show that the most significant results are obtained when the alarm time rate is rather large, within 0.25-0.45. For a comparison, they also simply use the same prediction algorithm but replace the detected anomalies in the magnetic signal by earthquake days themselves, so using seismicity itself as a predictor of future activity. This shows that the latter performs always worse, close to a random forecast. In order to investigate the influence of the parameters defining the position and size of the alarm window, they use the skilled area score of Zechar and Jordan (2008). From these computations, they deduce that several combinations of parameters may lead to an increased prediction power: the first one is characterized by a leading time of a week and an alarm window of less than four days; the other one by a leading time of 14-14 days and an alarm window of less than a week. The absolute optimum corresponds to a leading time of 8 days and an alarm window of 1 day, which seems to be amazingly precise for a prediction method. For those parameters, the Molchan curve is above the 95% confidence level when the alarm rate is within 0.10-0.57. However, the maximum probability gain is not very large, only 1.78 when the alarm rate is 5%. Indeed these results do not constitute a real prediction experiment, but rather the training phase of a potential prediction method. The optimal parameters provided by the authors should be validated by analyzing the success and failure rates on an independent dataset. Moreover, the results might change when considering the size of the area around the station (here, 100km).

### 3.4.3 Visible Spectrum (Earthquake Lights)

There have been numerous credible reports of luminous phenomena associated with earthquakes dating back for centuries, and high-quality images and video are now available due to the proliferation of digital imaging (Derr et al., 2014). Since light emission is necessarily associated with the motion of electric charge carriers, earthquakes lights provide conclusive and visible evidence of electromagnetic phenomena prior to and during earthquakes. Thériault et al. (2014) make a systematic study of these reports of earthquake lights and find that more than 90% of earthquake lights are reported near rift environments, marked by the prevalence of subvertic mafic dykes that have been enplaced during past periods of extensional tectonics.

Though early photographs of earthquake lights as described in Derr (1973) were met with skepticism, digital photography has now documented many instances. Precise timing of seismic waves and photography of earthquake lights in Peru has established that light emission is correlated with the arrival of seismic waves (Lira, 2008).

## 3.5 Thermal Infrared (TIR) anomalies

Eleftheriou et al. (2016) provide the only long-run experiment attempting to correlate earthquakes with TIR anomalies in Greece over the time period from May 2004 to December 2013, using the MeteoSat Second Generation–Spinning Enhanced Visible and Infrared Imager (MSG–SEVIRI). They consider 3151 TIR images acquired in the 9.80-11.80µm wavelength from 2am to 2:15am (local time). Data are first conditioned according to the month and time of the day of recording, as well as their type of geographical location (i.e. inland or offshore). This not only allows them to remove the background variations of temperature, but also to reduce the effect of other external parameters such as the vegetation cover. Anomalies are defined by comparing each conditional measurement to space and time averages, and their standard deviation. Note that all measurements affected by the presence of clouds are excluded from all those calculations. Side effects of the presence of such clouds (such as the cold spatial average effect, which results from the possible correlation of the clouds location with the ground temperature distribution) are also eliminated. Thermal anomalies are detected when the observed signal deviates more than four standard deviations from its reference value, and checked for spatial and temporal consistence (covering an area of at least 150km$^2$ within a 1°x1° cell, and occurring at least twice within a week).

This leads to the identification of 62 anomalies. A given anomaly is considered to be correlated with an earthquake with M≥4 if the latter occurs within a time window extending from 15 days before the anomaly to 30 days after it, and within the distance provided by the Dobrovolsky et al. (1979) law.

Results show an amazing correlation between anomalies and earthquakes, as 93% of the alarms correlate with seismic events, and only 7% constitute false alarms. However, their correlation procedures allow sequences of anomalies to occur (1) only before events, (2) only after them, or (3) to start before events and stop after they occurred. For events with M≥4, category (3) features less than 5% of the anomalies, while this share increases up to 15% in average for increasing magnitude thresholds. About 66% of the anomalies occur before events with M≥5.

In order to test the significance of their correlations, they use Molchan diagrams, which plot the fraction of missed events against the fraction of the space-time volume occupied by the alarms (which, in time, include 15 days before the anomaly and 30 days after it ended). They also consider a prevision mode, where they take only account of the 30 days following the anomaly. In both modes (correlation or prevision), they find that the observed associations have a significant gain over pure random ones. In correlation mode and M≥4, the gain is between 1.8 and 3.2, while in prevision mode it is between 1.5 (for M≥6) and 3.7 (for M≥5). In both cases, the largest gain is achieved when considering events with M≥5.5. In this case, the fraction of space-time filled with alarms is within 8-11%. Unfortunately, no study of the time distribution of anomalies is performed, and the earthquake catalog they use is not declustered beforehand. This latter feature may artificially increase their success rate.

One detailed example of a TIR anomaly prior to a major earthquake is found in the work of Piroddi et al. (2010, 2012, 2014). This study concerns the spatial and temporal distribution of TIR anomalies preceding the M=6.3 L'Aquila earthquake of April 06, 2009 in central Italy using the Nighttime Thermal Gradient (NTG) technique of Bryant et al. (2002, 2003, 2004a, 2004b). The NTG technique is based on the recognition that, under clear skies, the Earth's surface tends to cool during the night. This cooling trend can be obtained from geostationary weather satellites, which deliver calibrated TIR images every 15 min or 30 min. A linear regression of the implied radiation temperature versus time allows a slope $\Delta T$ to be derived, and this can be used to map nighttime temperature trends.

One notable result of this analysis is the fact that the recorded TIR anomalies are not present on the valley floor, where several active faults are located, including the Paganica Fault that produced the M=6.3 seismic event three nights later. The anomalies were associated with topographic highs on either side of the L'Aquila valley but not with the valley floor. This results suggests that the TIR anomalies are not due to warm gases emanating from the ground but rather the result of the accumulation of p-holes at topographic highs, where they undergo exothermal pairwise recombination to peroxy. The energy required to break the peroxy bond is on the order of 2.4 eV [*Kathrein and Freund*, 1983]. During pairwise p-hole recombination a fraction of this energy is regained, probably 2.1–2.2 eV. This energy will be deposited into the two oxygens participating in the recombination reaction, causing the m to becomes vibrationally highly excited to the tune of ≈20,000K equivalent. Hence, as new peroxy bonds form at the Earth surface, in particular at topographic highs, they are expected to emit infrared photons corresponding to the radiative de-excitation of the vibrationally excited peroxy bonds.

### 3.6 Ionospheric disturbances

Correlations between earthquakes and ionospheric disturbances have been first proposed after the occurrence of the Great Alaska earthquake in 1964 (Davies and Baker, 1965; Leonard and Barnes, 1965). Since then there have been many related publications (see for instance Cahyadi and Heki, 2013; Calais and Minster, 1995; Ho et al., 2013; Hsiao et al., 2009, 2010; Le et al., 2013; Lin, 2010, 2011, 2012, 2013a; Liperovskaya et al., 2006; Liu et al., 2001, 2006b, 2008, 2009, 2010a, 2011a, 2012; Mekela et al., 2011; Ondoh, 1998; Pulinets et al., 2003; Pulinets et al., 2004, 2005; Pulinets and Boyarchuk, 2004; Pulinets, 2007; Sarkar et al., 2011; Silina et al., 2001; Yao et al., 2012; Yu et al., 2009; Zhao et al., 2008, 2010; Zhou et al., 2009; Zhu et al., 2013a, 2013b). As for radon studies, most of this work has been devoted to case studies, such as for the Wenchuan (M7.9) event (Hsiao et al. 2010; Jhuang et al. 2010; Jin et al. 2011; Kamogawa et al. 2012; Lin et al. 2009; Lin 2013a; Liu et al. 2009, 2012; Xu et al. 2010; Yu et al. 2009; Zhao et al. 2008; Zhou et al. 2009; Zhu et al. 2013b), or the Tohoku (M9.0) event (Chen et al. 2011; Gokhberg et al. 2011; Hayakawa et al. 2013; Kamogawa et al. 2012; Le et al. 2013; Lin 2012; Liu et al., 2011a; Yao et al. 2012; Zhu et al. 2013b).

Ionospheric perturbations constitute the core of many studies (Chen et al., 2004; Depuev and Zelenova, 1996; Hayakawa, 2007; Hayakawa et al., 2006; Liu et al., 2006a,b; Liu et al, 2004a,b; Maekawa et al., 2006; Oyama et al., 2008; Pulinets et al., 2005; Singh, 2008; Trigunait et al., 2004; Zakharenkova et al., 2007) and generally focus on the fluctuations of the Total Electron Content (TEC), as well as on the changes in the Very Low Frequency (VLF) properties of the electromagnetic field. They rely on datasets acquired by satellites such as DEMETER (Detection of Electro-Magnetic Emissions Transmitted from Earthquake Regions), featuring instruments to measure directly the physical properties of the ionosphere. This micro-satellite of the French National Agency (CNES), weighting about 130kg, was a low altitude satellite launched in June 2004 and stopped operating in December 2010. It was orbiting at low altitude (710km at first, then 660km from December 2005), on a nearly polar (98° inclination) and nearly sun-synchroneous orbit, with 14 orbits a day and nearly 35min long half-orbits. The upgoing orbits corresponded to nighttime, and the downgoing orbits to daytime. The satellite did not strictly return above the same points every day, so that such consecutive 'nearby' points could be more than 1000km apart from one day to the next, and measurements did not cover latitudes beyond +/-65°. The initial mission of DEMETER has been to study the seismo-electromagnetic effects on the ionosphere prior to earthquakes, not to perform predictions *per se*. It featured three electric and three magnetic sensors, two Langmuir probes, an ion spectrometer, and an energetic particle analyzer. The frequency ranges covered by the magnetic and electric field measurements are respectively 10Hz-17kHz and DC-3.5MHz, while the sampling rate of the Langmuir probe for TEC estimation is 1s (Lebreton et al., 2006).

Some other satellites serving other purposes can also be used to infer the TEC. The Global Positioning System (GPS) consists of a set of 24 satellites, evenly distributed within 6 orbital planes, flying at an altitude of 20,200km. Each satellite emits signals at two distinct frequencies and carrier phases. Due to the dispersive properties of the ionosphere, each signal has thus a different speed. The carrier phase advance and group delay of those waves in the ionosphere depend on the electron density integrated along the full propagation path. The TEC can then be derived by comparing the phase delays between the two signals. A final correction is then applied to derive the VTEC, which corresponds to the TEC which would have been measured if the waves' rays had been vertical (Sardon et al., 1994; Leick, 1995; Liu et al., 1996). VTEC maps can then be derived every 30s (Tsai and Liu, 1999), using the more than 1,000 GPS ground-based receivers worldwide. Most of the works presented below and using GPS data use the global ionosphere maps provided with a 2h time resolution by the International GNSS Service (IGS), at distinct grid points with a spatial resolution of 2.5° in latitude and 5° in longitude, covering the full longitude range and latitudes within +/-87.5°. This provides maps consisting of 71x73=5,183 points each.

### 3.6.1 Total Electronic Content (TEC) measurements

In the last two decades, interest has grown in the estimation of the TEC in the ionosphere and its relationship with earthquakes (Astafyeva and Heki, 2011; Calais and Minster, 1995; Chavez et al., 2011; Kim and Hegai, 1999; Oyama et al., 2008).

*Measurements by DEMETER*

He et al. (2011) study the ionospheric electron density directly recorded by DEMETER, with a time resolution of 1s (Lebreton et al., 2006), and use data from 2006 to the beginning of 2009 as the satellite changed altitude late 2005. This reduces the dataset to 30,000 half-orbits, as they only keep nighttime periods to avoid solar perturbations. During the same time window, earthquakes amount to about 7,000 events with M≥5.0. A grid is centered on each event, featuring 11x11 cells with a 2° resolution. TEC data are segmented in 30s long intervals, which are then sorted in space within this grid. This allows them to compute, in each cell, a background average and standard deviation by considering data recorded from 31 to 75 days before the event and associated to a magnetic index Kp<2+. They then compute the average values in the last 30 days and Kp<3+ as the studied signal. This allows them to define a relative variation of the signal compared to the background (and normalized by the standard deviation of the latter) in each cell. This grid is then stacked and averaged over all considered events. This allows them to show evidence for a maximum anomaly close to the epicenter, located slightly to its North in the Northern hemisphere, and slightly to its South in the Southern hemisphere, while its spatial extent is about 350km. The anomaly is also more pronounced for offshore events, for larger magnitudes and shallower depths. Removing all data that follow an event within a day (to eliminate the effect of gravity waves possibly triggered by the latter), the effect is weaker but clearly identifiable. Looking more closely at their results, we can notice that the reported anomaly is positive, but that its normalized value (by the background's standard deviation) is only 0.68 at maximum, i.e. of about one standard deviation. We also notice that such a systematic shift between the event and the anomaly has also been documented for Radon anomalies (Hauksson, 1981), but He et al. (2011) did not look at any possible relationship between the amplitude of the shift and the size of the associated event.

Ryu et al. (2016) compute the equatorial ionization anomaly (EIA) and study its relationship with mid-latitude seismic activity. This is done by selecting a restricted zone in North-East Asia (roughly covering a 40° by 20° window centered on Japan) and earthquakes with magnitude M≥6.0 occurring therein during the DEMETER mission (35 events in all, previously part of the analysis of Liu et al., 2013). The quantity they consider is, for each orbit, the TEC measured near the geomagnetic equator region, by estimating the normalized equatorial plasma density (NEDP) which is ratio of electron density within 15° of the geomagnetic equator to the same averaged value at latitudes between +/-30° and +/-50°. Their analysis is based on plots of the TEC as a function of latitude for profiles distant to a given epicenter by less than 20° in longitude. They first perform a restricted set of individual case studies of 7 events with M≥6.8 in a time period extending from one month before to one month after each selected event. In a few cases, they outline that some anomalies could be due to the occurrence of other large events occurring outside of their spatial window (for instance in Solomon Islands, Taiwan, China, etc...), an hypothesis that is difficult to test. They finally claim that 5 out of the 7 events are associated with equatorial positive TEC anomalies within 10 days before their occurrence. Looking at individual TEC profiles for the M≥6.0 events, they do not observe such an increase. Anomalies appear more clearly if one selects only M≥6.5 events (i.e. 16 events in all, with only one not showing such an increase, which can be explained by its larger focal depth). The effect seems to be absent for hypocentral depths larger than 30km. They conclude that mid-latitude seismicity affects the equatorial ionosphere in the morning.

### *Measurements by GPS*

Saroso et al. (2008) study correlations between TEC and earthquakes occurring in Sulawesi during the time period 1993-2002, using three ground-based GPS stations. The TEC is first estimated at the location of given events, and for any local time. The mean and standard deviation of this distribution are estimated, as well as upper/lower detection thresholds (corresponding to 1.34 standard deviations above or below the mean, i.e. a 82% confidence level). This reference state is then compared to TEC data derived at the same locations each of the 15 days preceding an earthquake, and allows one to infer that anomalies occur within 2-7 days before the events. However, the earthquake dataset which is chosen by the authors seems to be arbitrary, as they limit it to 11 seismic events with M≥5.9, but do not mention how they selected them, as well as to the Sumatra, 2004 event (M=9.3). They also do not mention any declustering of the catalog, so that some anomalies may be each related to several distinct events occurring as clusters (thus increasing the apparent success rate), and they do not test any null hypothesis (such as the distribution of anomalies within similar 15 days time windows in aseismic areas). A last criticism is that they do not further test whether some of the anomalies could correspond to some anomalous geomagnetic storms or other external disturbances.

The studies that follow have been conducted at a worlwide scale using the TEC maps provided by the IGS and described above. Le et al. (2011) study TEC anomalies at the global scale before events with magnitude larger than 6, within the time period 2002-2010. Earthquakes occurring within 4 days after a magnetic storm are excluded from the analysis, as well as those occurring at nearly the same location and within 15 days in time, leaving a final and significant set of 736 events. The VTEC is interpolated linearly in time with a 1h step, before modeling them spatially using a spherical harmonics expansion. The grid point that is the closest to each event is considered as its associated TEC data point. At each time sample, they compute the mean and standard deviation of the signal within the 1 to 15 previous days, and the current point is considered as an outlier if it is beyond a single standard deviation from the mean. If there are more than 6 successive anomalous data within a given day, they associate to the latter the largest quantile reached by such a fluctuation (R=60%, 80% or 100%). Each of the 1-21 days preceding each selected earthquake is checked for its abnormality and quantile level R. The days with anomalous geomagnetic indices are excluded from the statistics, as well each of the three days following them. The results show that: (i) the rate of observed anomalous days increases with the magnitude of earthquakes for shallow depths (≤20km); (ii) the rate of anomalies is larger when time gets closer to the occurrence of the event; (iii) positive anomalies are observed twice as often as negative ones; (iv) the rate of anomalies decreases with the focal depth. No specific variation with latitude is observed. In order to check with a null hypothesis, the 61-300 days before each earthquake are considered as background days. If another earthquake associated to the same node occurred within these background days, the 15 days before it and after it are removed, as well as anomalous geomagnetic indices days and their three following days. The background rate of anomalous days is 2 to 4 times smaller than the rate of anomalies in the 1-2 days before the events, confirming the significance of the observed precursory signal.

Yao et al. (2012) provide a worldwide analysis of M≥7.0 events during the year 2010, considering only those that have not been preceded by another M≥7 event within a 15 days window (yielding a final set of seven events in all). The TEC maps are then linearly interpolated at the location of each event. A sliding window method is applied, allowing at any given time to compute the mean and standard deviation of the signal over a 30 days window before that time. Thresholds are defined at +/-1.5 standard deviations from the local mean (corresponding to a confidence level of 87%). They then check for TEC anomalies within 0-14 days before each event. This analysis is repeatedly performed at each node of the original data grid. While some anomalies are clearly correlated with external perturbations, some others are not. Using only the latter, they check that 5 out of 7 events are preceded by TEC anomalies, which may be either positive (4 cases) or negative (1 case). Anomalies occur during daytime (from 12:00 to 20:00, local time), within 2 days before the events. By looking at anomalies at all the grid points, they notice that they indeed occur all over the world. Yet, some anomalies repeat in time at some of the nodes, and the highest repeating rates tend to cluster spatially in regions close to the future events, even if the spatial mode of the anomalies' density does not coincide with the epicenter (yet no systematic shift is observed). In parallel, some weaker effects are observed in the magnetically conjugated region, while the strength of the effect does not seem to be correlated with the depth of events. Unfortunately, the authors offer no such analysis and results over locations and times with no earthquake activity.

Zhu et al. (2014) study the anomalies before M≥7.0 earthquakes worldwide during 2003-2012, which amount to 144 events in all (as events occurring within a 5° distance and within less than 15 days in time are excluded from the analysis). The nearest grid point is used as a proxy for each event for the TEC signal. They use a sliding window of 10 days before any given time to compute the mean and standard deviation of the TEC signal at that time. They apply this technique to each of the 15 days before each event, and consider a confidence level of 95% to detect outliers. They exclude days when magnetic disturbances occur, as well as their two following days. Computing the relative amount of events before which some anomalies have been observed, they observe a larger rate of anomalies when the magnitude increases, and that the rate of negative anomalies increases as the time of the event approaches. Yet, the effect is very weak as the increase is of the order of a few percent when M varies from 7.0 to 7.9. They also observe that most negative anomalies occur between 12:00 LT and 18:00 LT while their duration is about 2 hours. No such specific pattern is reported for positive anomalies, and this effect seems to hold only for events with M≥7.6.

Ke et al. (2016) look at GPS-derived TEC anomalies before M>5.0 events in China from 2003 to 2013. They have unfortunately chosen arbitrarily only 24 events (as we can check that this set of events does not obey the Gutenberg-Richter distribution). Similarly to Yao et al. (2012), they use IGS maps published every 2h. They also use a sliding windows approach, but the chosen parameters are not indicated clearly, yet seem to be apparently the same as in Liu et al. (2010a,b), i.e. a time window of 5 days. Anomalies are defined by first computing the upper and lower quartiles of the TEC data within the 15 days before each event, and by defining upper and lower thresholds for detection by the quartiles-to-median distance multiplied by 1.5. Looking at anomalies from 15 days before to 5 days after the events, they observe that they can be positive or negative, occur before or after the events, and do not depend on earthquake magnitude. For 5<M<5.9, 33% of events have only positive anomalies, while 42% of events have only negative anomalies. The remaining 8% have both positive or negative anomalies. For 6.0<M<6.9, those ratios reach respectively 44%, 33% and 11%. For 7.0<M<8.0, they are 50%, 25% and 25%. Drawing spatial maps of TEC anomalies at the time of the events, they show that anomalies occur close to the events, but also at more distant locations. Unfortunately, no quantitative test of the significance of the results is provided. They also show that the amplitude of the anomalies fluctuates a lot within the 5 days before the events, and tend to be more uniform afterwards (but are not smaller, despite their claim). They finally note that more extreme solar and geomagnetic activities occur before most M>7 events, which they interpret by claiming (!) that these geomagnetic perturbations are indeed triggering those seismic events.

### 3.6.2 Ion density

The DEMETER satellite is also equipped with an instrument labelled IAP (Instrument Analyseur de Plasma; see Berthelier et al. , 2006) allowing to estimate the total ion density (i.e. the sum of $H^+$, $He^+$ and $O^+$ ions ), with a 4s time resolution in survey mode and twice that sampling rate in the burst mode. This allows them to get an alternative measurement of the TEC as both signals are assumed to fluctuate in opposite ways. The advantage of using ion density is that it is a less noisy signal than electron density, thus allowing to detect anomalies more efficiently. At the DEMETER altitude, the ion density is dominated by the $O^+$ content (Li and Parrot, 2013).

Parrot (2011) analyzes a limited dataset extending from August 2004 to October 2009. They consider seismic events with M≥4.8 (17,366 earthquakes in all) and the measured ion density from 0 to 15 days before the events, keeping only the closest nighttime orbit within 1500km of each epicenter (which corresponds to a flight time of about 3 minutes). Sometimes, there is no such orbit within a given day (as the satellite is too far or recording failed). Anomalies are detected by a simple detection of local maxima of the ion density, and their amplitude is normalized by the background value along the same orbit segment. They compare their results to two reference datasets: the first one is obtained by switching the latitudes and longitudes of earthquakes but keeping their times (RAND1). Another dataset is obtained by shifting their longitude 25° to the West (RAND2), in order to keep the latitudinal structure of seismicity intact, as most events occur close to the Equator and as there are more anomalies of external origin around the Equator during nighttime. Such surrogate catalogs are designed to provide a background rate of spurious associations between anomalies and earthquakes. If *N* is the number of days between an event and an aftershock, and if *N*<15days, only ion density data within the *N*-1 days before the aftershock are used (but the aftershock is not eliminated from the catalog). For events occurring below the sea level, the amount of detected perturbations associated to events is always larger than the background rate. Inland, similarly high rates are observed only when considering M>6 events. Considering the average amplitudes observed before events, as well as the average maximum amplitude, both are found to be larger for the natural catalog, increase with the magnitude of the event, and are larger for offshore events (yet, no error bars are provided). This latter result is interpreted as the existence of a larger electric conductivity above the sea.

Parrot (2012) repeats the same analysis, but now splits events into three distinct categories: below the sea level, inland, and close to the coast, retaining only the closest orbit to each event, and eliminating data when $K_p$>3+. He then considers the median of the largest anomaly observed in the 15 days before each event. The observed anomalous effect increases with magnitude, decreases with focal depth, and is again stronger offshore.

Li and Parrot (2012) average consecutive values in the burst mode to get the same rate as in the survey mode. The data are then smoothed using a Savitzky-Golay algorithm (Savitzky and Golay, 1964). Anomalies are detected using the same methodology as above, with durations constrained to be between 23s and 2 minutes (corresponding to flight distances of up to 840km and at least 5 sampled points). Events are split according to their magnitude (M4.8-5.0; M5.1-6.0; M≥6.1) and focal depth (larger and smaller than 20km). Anomalies are then associated (or not) to earthquakes according to the delay times T between them (<7 days or <15 days), and the epicenter-anomaly distance D (<500km, <1000km, or <1500km). For D<500km and T<7 days, the number of false alarms is 44,863 over a total of 46,446 i.e. 97%. The rate of good detections increases from 14% to 18% when the magnitude increases (see Tables 2 to 7) and decreases slightly with depth. Logically, it increases when cutoffs on time interval and distance increase. Unfortunately, the rate of false alarms is not systematically indicated and they do not provide the total amount of positive anomalies, in order to compare with the ones successfully associated to events.

Using a technique similar to Li and Parrot (2012), Li and Parrot (2013) present results about ion density variations, focusing on nighttime anomalies only, resulting in 96,863 half-orbits, with 27,257,933 samples. Different criteria are applied to define an anomaly: (i) a duration must be between 23 and 120s; (ii) the distance from the corresponding satellite position to the nearest seismic zone must be less than 1,500km (considering only events with M>4.8 during the satellite mission as defining seismic areas); (iii) on the day of the anomaly, $K_p$ must be smaller than 3 in order to remove the effect of solar activity. Spurious peaks are also removed, without mentioning the criteria for their detection. The anomaly time window is defined by a local extremum bracketed by the two closest points where the derivative of the signal changes sign, while the values of the signal at the end points of this window define the local background signal. This allows them to present evidence for 56,139 such anomalies (which are observed to be either positive or negative), while the total number of earthquakes is 21,863 (using the USGS catalog from August 20, 2004 to December 31, 2010). Computing the ratio $A$ between the anomaly's amplitude and the associated background value allows them to deal safely with problems such as a change of the satellite altitude. The comparison with earthquake activity is performed by computing the distance $D$ between an anomaly and an event (with a maximum value of 1500km), the time delay $T$ (considering a maximum value of 15 days), and the depth $d$ of the event. If an earthquake can be associated to one or more anomalies, it is counted as a single detection. If not, it is a bad detection. If a perturbation corresponds to an event, this is a true alarm; if not, a false alarm.

Considering $A>10\%$ and targeting events with M=4.8 to 5, the number of false alarms amounts to 63% (26,877 alarms in all). The number of wrong detections is 64% (over 12,057events). If we take account of earthquakes with magnitude larger than 5, the false alarm rate drops to 19%. Counting all events with M>4.8, the rate of false alarms drops from 23% to 17% when $A$ increase from 0 to 15%, but the rate of bad detections then increases. Anyway, the rate of good detection increases with the magnitude of events, and about 90% of the right alarms correspond to upward (positive) anomalies. The number of associated anomalies per earthquake is also observed to increase with their magnitude, while the amplitude of the anomalies is only weakly correlated with the magnitude of the events.

They then consider 3 types of events: inland, below the sea with a water depth larger than 1km, and below the sea but close to the coast. The percentage of good detections is larger under the sea and is the worst when close to the coast, confirming the results of Parrot (2012). Unfortunately, they do not mention anything about the rate of such correct alarms. In order to compare the results with a reference dataset, they shift events 25° Westward, and 1 month backward in the past (in order to keep intact the latitudinal structure). The ratio of good detections is then 42,27% (with a standard deviation of 0.39), whereas the ratio is about 44.7% at worst with the true data. The observed effect is thus very weak. They also find that one perturbation is at most associated with a single event, while each event can be associated with more than one perturbation. Intracontinental seismicity features only events with less than 9 anomalies. Events with 10-19 anomalies occur mostly at plate boundaries, while events with more than 20 anomalies occur in a very specific zone in the southern hemisphere.

In order to complement their analysis, they also look at the number of perturbations before large events as a function of time. The total number of perturbations before events reaches 64% during the week before (78% over the Himalayas), while no such increase is observed in the southern zone mentioned above. The number of anomalies shows a maximum on the day of the event, and decreases for larger time delays. Results using electron density are about the same, except that the peaks are a bit less sharp. Both the rates of false alarms and good detections decrease when small anomalies are eliminated. As the satellite flies over any zone only a few minutes a day, chances are large to miss them (and have a wrong detection) if the anomalies do not occur continuously. They also notice (focusing on only two events of M=8.8 and 6.3) that the epicenter lies close to the barycenter of the associated anomalies. However, this could be due to the uniform random distribution of anomalies, which are constrained to sample a symmetric spatial zone around the earthquakes.

### 3.6.3 Electric and magnetic fields measurements

*Electromagnetic perturbations*

The DEMETER satellite was equipped to measure simultaneously the electric and magnetic fields using respectively the ICE (Instrument Champ Electrique, see Berthelier et al., 2006) and the IMSC (Parrot et al., 2006; Santolik et al., 2006). The ICE used two distinct modes of acquisition: (i) a survey mode, to record low bit rate data at 25kbits/s; (ii) a burst mode, triggered above predefined seismic regions, to record high bit rate data at 1.6Mbits/s. The seismic regions featured mainly the Southern American subduction zone, the Alps-Zagros-Himalaya mountain ranges, parts of the Asian pacific subduction zone, and a couple of other areas. The ICE featured four antennas to record the three components of the electric field. This analog signal was filtered into four frequency channels, then digitized and stored:

- DC/ULF (0-15Hz): the waveforms of the four measured potentials, digitized at 39Hz and stored for both modes of operation.
- ELF (15Hz-1kHz): three components of the field waveforms, sampled at 2.5kHz, in burst mode only.
- VLF (15Hz-17.4kHz): one component of the field waveform, or spectral data for one of the components, are sampled at 40kHz. In burst mode, the power spectrum is computed at a resolution of 19Hz. Forty such spectra are averaged and normalized, using a resolution of 2s, and both the waveform and spectra are stored. In survey mode, only the spectra are stored, according to three submodes: (0) identical to the burst mode; (1) the temporal resolution is 0.5s; (2) the time resolution is 2s but the frequency resolution is 78Hz.

- HF (10kHz-3.175MHz): the same field component as in the VLF channel is sampled at 6.66Mhz. Averaged power spectra are computed every 2s with a resolution of 3.25kHz over 40 intervals of 0.6ms. When in burst mode, the average spectrum is stored as well as the waveform of one of those intervals. In survey mode, only the average spectra are stored. There are three submodes: (0) power spectra with a resolution of 3.25kHz ; (1) same, but with a time resolution of 0.5s; (2) the time resolution is 2s and frequency resolution is 13kHz.

Nemec et al. (2008) use about 2.5 years of data, focusing on VLF band. The power spectra of one electric and one magnetic component are computed onboard with a 19.5Hz resolution in frequency, and a 2s or 0.5s resolution in time. They use the electric component normal to the plane of the orbit, whereas the magnetic one is inclined 45° from the satellite velocity vector. Their analysis considers 11,500 hours corresponding to 20,000 half-orbits. The corresponding USGS catalog lists about 9000 events with M≥4.8. The dataset is first partitioned into a 6-dimensional matrix, defined by: geomagnetic longitude (res. 10°) and latitude (2°), frequency (16 bands for the electric component and 13 for the magnetic one, of 117Hz each, smaller than 10kHz in order to avoid the influence of terrestrial VLF transmitters), $K_p$ index (0 to 1o, 1+ to 2+, and >3), magnetic local time, and season (October-April and May-September).

Within each cell of this matrix, they estimate the empirical cumulative and density distributions of the observed power spectrum amplitudes. For each considered observation $E_i$ within a cell, the corresponding cumulative probability $F_i$ can then be computed. Such $F_i$ values are sampled only when the position of the satellite flies close to an earthquake epicenter (i.e. within 1,100km and up to 5 days before and 3 days after it occurred). Such datapoints are removed if some seismic events are too clustered, unfortunately without any mention of the rejection criteria in space and time. The sampled values of $F_i$ are then binned as a function of frequency range (using the same bins as defined above), time to or from an event (using a 4h resolution), and distance to the event (using a 1°=110km resolution). The probabilistic intensity $I_b$ is then defined within each bin, as the average value of $F_i$ minus 0.5. Applying a correction to take into account the fact that all measurements are not independent, they detect anomalies of $I_b$ beyond 3 standard deviations. The same method is applied on two types of surrogate catalogs: either keeping the real locations of earthquakes with random occurrence times, or keeping the real occurrence times with random locations. No specific pattern of $I_b$ is noticed when using any of the random catalogs. The original data are then split according to day/night, and according to the focal depth of earthquakes (<40km and beyond 40km). Keeping only data recorded at night, they show that $I_b$ decreases by about 3 standard deviations within a time window extending from 0 to 4 hours before seismic events with M>4.8. Using only superficial events with M>5.0, the observed decrease is about 4 standard deviations. The latter corresponds to a 4 to 6dB decrease of the power spectrum. The effect is weaker when using magnetic data, while no effect is observed during daytime or for deeper events. The effect is also observed to be stronger within a spatial window of about 350km, which corresponds well with similar estimations using ion density. Their interpretation is that during daytime, ionospheric ionization is so large that it masks all amplitude changes due to seismic activity. They argue that the 1.7kHz frequency corresponds to the cut-off frequency of the earth-ionosphere waveguide during the night (Budden, 1961), the low frequency cutoff being inversely proportional to the height of the ionosphere. Nemec et al. (2010) argue that the source of VLF radiation recorded by DEMETER during nighttime is due to electromagnetic waves generated by thunderstorm activity. If the height of the ionosphere decreases, then the cutoff frequency increases, and the power spectral density at 1.7kHz decreases.

Nemec et al. (2009), analyze the same dataset using 3.5 years of data (corresponding to 9,000 hours in nighttime for about 15,500 orbits), during which 9,500 earthquakes have been recorded, with a depth smaller than 40km and M≥4.8, and 5,500 events with M≥5.0. They first focus on the 200Hz-wide frequency band around 1.7kHz, nighttime recordings, M>=5.0 and depth <40km events, and a satellite-epicenter distance smaller than 3° (about 330km). Plotting $I_b$ as a function of time, they still observe a sudden decrease beyond 3 standard deviations before the events. But the decrease is now only 2.4dB to 3.6dB, smaller than observed by Nemec et al. (2008). They interpret this change of amplitude as maybe due to some scarce but large contributions of the background level. Interestingly, they quantify the usual natural background variability of the raw signal to be about 11dB, which is much larger than the observed anomalies. The latter are indeed detected only because of their stacking procedure, but such anomalies could certainly not be detected individually when running, for instance, a genuine prediction experiment.

In a second analysis, they use a new data processing in order to study spatial patterns. They then re-estimate the $F_i$ values taking account of a single time bin (0-4 hours before events) in order to infer the spatial location of anomalies relative to the observed earthquakes. To do so, they consider any point P within 10° in latitude and longitude from a seismic event. For each orbit passing within 3° of such a point P, they check if the values of $F_i$ are larger or smaller than usual. This is done by using the non-parametric Mann-Whitney U test, which tests if two populations have the same mean (see Sheskin, 2000), with a confidence level of 0.01 (their results being insensitive to this arbitrary choice). This allows them to compare the observed distribution of $F_i$ values close to point P to their distribution along the rest of the same half-orbit, considered as a background signal. They then test if the observed Fi's are smaller, larger or indistinguishable from the values on the rest of the half-orbit, the latter cases being eliminated from the rest of the analysis, so that they only study cases where an anomaly is observed before an event. They then test the number of increases and decreases using a simple binomial distribution and assuming that both rates should not differ significantly if no effect holds. Despite the strong limitations of this test (as they exclude cases where no anomaly is observed before an event), they find a nonrandom, anomalous pattern about 1.5° North and 2° West of seismic events when stacking over all seismic events. The probability of random occurrences of increases and decreases of $F_i$ is about $10^{-4}$ at this location. They finally focus on the most anomalous point (and its 3° surroundings) to check how the numbers of increases and decreases depend on earthquake magnitude, depth and altitude of the solid surface above the event. For M>5.5, there is most often a decrease, the pattern becoming more random for lower magnitude events. An observed decrease is also more probable for depths <20km. No dependence is found on the inland or offshore location of events. In a tentative explanation of the Westward drift of the anomaly relative to the earthquakes, they suggest that either aerosols are subjected to the Coriolis force, or that some ions are subjected to a magnetic field. But they offer no reason for the northward component of the drift. We can mention a similar shift of the anomaly on Radon anomalies recorded on the ground. In terms of statistics, it should be noted that, for 50% of events, no deviation occurs according to the Mann-Whitney test.

Pisa et al. (2012) use the same method as Nemec et al. (2008), now applied to the full DEMETER dataset. The data are initially sorted within the same 6-dimensional matrix. Data that can be related to more than one seismic event are removed. They focus on nighttime half-orbits, seismic events with M≥5.0 and depth ≤40km. They still observe a decrease of power within the same frequency band, with the strongest significance when distances are smaller than 440km, and still within the 4 hours window before events. The decrease is now about 3 standard deviations, which corresponds to about 1.8dB, while the natural background variation is estimated to about 7.5dB. The effect thus seems smaller as data accumulate. Yet, the probability to be a random positive/negative deviation is still only about 0.1%, using a t-test or a binomial test.

Pisa et al. (2013) provide additional observations using the survey mode during the whole mission. They consider different earthquake magnitude ranges (from M≥5, about 12,000 events, down to M≥3, with about 153,000 events in all). They still consider the 6-dimensional matrix to condition the data, but the satellite-epicenter distance varies from 0 to 440km, and the time interval from 2 days before to 1 day after each event. They also use quarters instead of semesters for conditioning on the season. Removing cases where data can be associated to more than one event, they are left with 8,400 events. They observe the same result as before: an attenuation of 2dB in the last 4 hours and within 440km, which is about 2.9 standard deviations (with a probability to be random of about 0.3%). In all other distance-time bins, the probability is found to be one to two orders of magnitude larger. Two surrogate catalogs are then submitted to the same analysis: one in which seismic events are shifted 10 days back in time, and another one where they are spatially shifted 25° to the West. No signal is observed for both surrogate catalogs. In the case of the natural catalog, the effect is much stronger between March and August, which is the season of lightning (Christian et al., 2003). It is also slightly stronger for larger latitudes (>20°), and for events below the sea surface. This is interpreted by the lower attenuation of the VLF waves in the Earth-ionosphere waveguide above the sea (Meyer et al., 2011), as the main contributing factor is the conductivity of the surface of the lower boundary of the waveguide. The effect is also stronger for events with shallow depth (≤33km), but many events are attributed an arbitrary depth at 10km (which happens when the location algorithm fails to constrain depth). A more refined temporal analysis shows that the decrease is more pronounced 3 to 4h before the events, with an amplitude of about 2.3dB. The effect increases with the magnitude of the events and is significant when M≥4. As pointed out by Toledo-Redondo et al. (2012), the height of the lower boundary of the ionosphere exhibits a seasonal variation and depends on the position on Earth. Pisa et al. (2013) then claim that the electric conductivity of the lower troposphere increases, because of charge carriers emanating from stressed rocks before major events. The air conductivity increases, the electric current down from the ionosphere also increases, by Ohm's law (Rycroft and Harrison, 2012), so that the ionosphere lowers at night time at about 88km, by about 2km (Harrison et al., 2010). They propose that VLF radio waves originating from lightning and propagating in the Earth-ionosphere waveguide at night are cut-off at a slightly higher frequency (1.74kHz vs 1.7kHz): there is thus additional attenuation of signals from lightning propagating in this waveguide.

## IV-Discussion and conclusion

This review provides a heterogenous but encouraging assessment of the correlations between many non-seismic signals with tectonic and earthquake activity, which are expected to occur according to the theory based on the activation of peroxy defects and flow of the associated p-holes.

To recapitulate, peroxy defects are point defects in the matrix of oxide and silicate materials, including the overwhelming majority of rock-forming minerals, in which pairs of oxygen anions have converted from the usual 2– valence state to the 1– valence state, forming a very short $O^-–O^-$ bond. Though peroxy concentrations have not yet been systematically studied, there is strong evidence that peroxy defects of the type $O_3$X-OO-Y$O_3$ (with X and Y standing for $Si^{4+}$ or $Al^{3+}$ etc.) are ubiquitous in the constituent minerals of igneous and high-grade metamorphic rocks. The reason is that peroxy defects derive from hydroxyl "impurities", commonly $O_3$(X,Y)-OH, that become incorporated into the matrix of any minerals that crystallize from an $H_2O$-laden melt or magma or that recrystallize in any high temperature $H_2O$-laden metamorphic environment [*Friedemann T. Freund and Freund*, 2015]. The formative reaction involves a redox conversion of solute hydroxyl pairs as described by eq [1], whereby electrons rearrange in such a way that the two hydroxyl protons, $H^+$, take over an electron from their respective hydroxyl oxygens, $O^{2-}$, changing into two H which combine to form $H_2$, while the two $O^{2-}$ change to $O^-$, which form a peroxy bond, $O^-$–$O^-$. This is a classical redox reaction that causes one part to become reduced and the other to become oxidized.

As long as the peroxy bond is intact, it is electrically inactive. When the peroxy bond is perturbed, for instance by bending during the application of mechanical stress, it breaks. In the process described by eq [2] the peroxy bond becomes unstable and takes over an electron, e', from a neighboring $O^{2-}$. The neighboring $O^{2-}$ thereby changes into $O^-$, a defect electron in the oxygen anion sublattice, hence a hole. Because of its unusual transport properties, foremost its ability to spread out fast and far, this type of charge carrier has been called a "positive hole".

Many of these independent studies provide results that seem to be consistent with the prediction of the theory, even if some differences blur a bit the full picture.

First, even very tiny stress fluctuations in natural settings seem to possess a non-seismic signature, as revealed by Radon analysis and, to a somewhat lesser degree, electromagnetic fields. This is indeed compatible with the empirical relationship provided by Dobrovolsky et al. (1979), suggesting that strains as low as $10^{-8}$ may be associated with such signals. This provides a strong tie with all non-destructive laboratory experiments conducted under well-controlled mechanical and physical conditions. For instance, Scoville et al. (2015) report on rock stressing experiments changing the stress rate over 8 orders of magnitude. One of the most remarkable observation is that, when a fine-grained gabbro is subjected to stress, the outflow of p-holes from the stressed subvolume is extremely sensitive to very low stress level changes. A plausible explanation is that many peroxy bonds exist at grain boundaries and across grain boundaries. Ever so slight shifting of mineral grains relative to each other, such as during small stress changes, make these peroxy bonds highly susceptible to dissociation and instant release of highly mobile p-holes.

The same paper [*John Scoville et al.*, 2015] also reports on the lifetimes of stress-activated p-hole charge carriers. It shows that, inside the stressed subvolume of the fine-grained gabbro rock, the p-hole lifetimes spread from milliseconds to several months. It is therefore quite conceivable that weak mechanical forces such as experienced during the tides will suffice to reversibly activate a fairly large number of p-hole charges in the deep crust, thereby providing an explanation for the diurnal and semi-diurnal patterns recognized, for instance, in radon release data (Teng and McElrath, 1977). The concept of peroxy bond breakage during grain-grain sliding is further supported by the observation that, when the stress is removed, allowing the grains to return to their starting positions, the p-hole charge carriers quickly recombine and return to their inactive peroxy state. However, the nature of the relationship between the size of the Dobrovolsky precursor zone and the magnitude of the upcoming events is not yet clearly established.

Combining a wide array of studies clearly indicate that earthquakes are often preceded by various anomalies a few days to weeks before they occur. Yet, there is no one-to-one relation between signal anomalies and earthquake nucleation. What most of pubished studies show is that, based on a single precursory indicator, the number of false alarms is large: anomalies can and do often occur without subsequent seismic event. However, it also appears that the rate and intensity at which anomalies are recorded increase as the time of the event approaches and as its magnitude increases. This statement holds even though, in many cases, the amplitude of any given precursory anomaly is not significantly correlated with the magnitude of the seismic event. The leading time of the anomalies, as well as their duration, also seem to correlate positively with the magnitude of the earthquake.

One complicating factor is that many studies report an offset between the spatial location of the anomalies and that of the future epicenter, often a few hundreds of kilometers. There are potentialy several mechanisms that could lead to such a shift. First, just as currents in the ground can induce changes in the ionosphere, it is well known that currents in the ionosphere or magnetosphere (or many other currents) can induce telluric currents in the ground. Several drivers of telluric currents are described in Helman (2013). Atmospheric precursors can be driven either by ionospheric currents, in the case of ionized species, or by winds, Coriolis forces, etc, in the case of neutral species. It is also possible that stress buildup may occur somewhere along a fault, leading to observed precursors, while rupture may occur somewhere else along the fault. A careful examination of more radon data is warranted, especially as the spatial density of continuously recording stations has increased since the compilation paper of Hauksson (1981).

Global analyses, which can only be achieved by the use of measurements by satellites, allowed for the most systematic analyses (typically at the scale of years to a decade) that can be compared with significant earthquake catalogs featuring tens to hundreds of thousands of events. Most of these works show that precursory anomalies tend to be more significant for larger magnitude events, when the focal depth is smaller, and when events are associated with offshore subduction zones. They diverge on the time at which such observations can be made. For instance, ionospheric perturbations deduced from GPS measurements suggest that TEC anomalies at the geomagnetic equator occur in the morning for events in the afternoon. Most TEC anomalies are reported to be negative, while ion content anomalies (as reported by the DEMETER satellite) tend to be positive, which is perfectly consistent. However, different satellite technologies come to different conclusions about specific topics: for instance, the DEMETER daytime data are systematically eliminated as they do not display specific patterns of anomalies, which is interpreted as due to noise arising from external influences. This finding questions the anomalies patterns found using GPS data, which provide evidence for anomalies only during day time. It also seems that, when using the DEMETER dataset for ion densities, the strength of anomalies seems to depend on whether the earthquake focus is offshore or inland. When analyzing data derived from the electron distribution this relation is not found.

The various methodologies used to detect anomalous behaviors are mostly local in time, meaning that the signal observed at a given time is compared to a background value that is estimated up to at most a few weeks before. In contrast, a large part of the anomaly detection procedure using the DEMETER dataset estimates the statistics of the background signal using the whole dataset at hand. This last procedure can be prone to mistakes if the background signal itself is not stationary in time. As a result, increasing the size of the dataset would lead to fewer detected anomalies. This might explain why the electric field anomalies recorded in the VLF range are observed with smaller and smaller amplitudes as the size of the considered dataset increases. Indeed, preliminary results (Kamer et al., work in progress) indicate that the background signal is not stationary, even in aseismic areas such as Bostwana, for instance.

Another important aspect that has been neglected so far is the temporal and spatial clustering of earthquakes. Some works take this universal process into account in order to remove seismic events occurring too close in time in order to eliminate double counting of successes (or failures) when correlating them with anomalies. Unfortunately, all the techniques used for this declustering are somewhat primitive, if not arbitrary. One of the goals of statistical seismology is to model seismic catalogs with sets of well defined clusters whose seeds are distributed randomly in time. This allows one to compute the probability of any event to be either independent or to have been triggered by the cumulative effect of some previous events. Preprocessing seismic catalogs with the latest generation of such sophisticated declustering tools (Nandan et al., 2017) should certainly help in assigning probabilities of each detected anomaly to be associated to any observed earthquake.

Finally, a striking common feature of all these works is the absence of any assumption about the morphological features of the anomalies we should look for. In a sense, this allows one to be as objective and open minded as possible when looking for them. But this is also a severe drawback, as the probability is certainly high to mistake transient noise for anomalies. Some recent works suggest that anomalies in the magnetic field recorded by ground stations have the shape of unipolar pulses, an observation confirmed by experiments and numerical simulations. The latter either solve differential equations at the microscopic level (Scoville et al., 2015), or use a coarse-grained description of a frictional fault as an assembly of blocks separated by elastic springs (in the spirit of Burridge and Knopoff, 1967), each unit of the model being also modeled as a RLC circuit within which charge generation in each block depends on the applied stress upon it (Chen et al., 2017). These works need to be developed further in order to provide guidance concerning the various transfer functions that would help modelling signatures of seismic-induced anomalies.

# V-References